\documentclass[prb,twocolumn,aps,longbibliography,superscriptaddress,notoc,10pt,floatfix]{revtex4-2}
\usepackage{bm}
\usepackage{graphicx}
\usepackage{amssymb}
\usepackage{amsmath}
\usepackage{color}
\usepackage[utf8]{inputenc}
\usepackage{ulem}
\usepackage[unicode=true,colorlinks=true,citecolor=blue,urlcolor=blue]{hyperref}
\usepackage{orcidlink}
\usepackage{tabularx}

\graphicspath{{}}

\newcommand{\nix}[1]{}

\def \CTS{Co$_{1/3}$TaS$_2$}

\def \TNo{$T_{\mathrm N1}$}
\def \TNt{$T_{\mathrm N2}$}
\def \Tb{$T<T_{\mathrm{N2}}$}
\def \Ti{$T_{\mathrm{N2}}<T<T_{\mathrm{N1}}$}
\def \Th{$T>T_{\mathrm{N1}}$}

\begin{document}

\author{Erik Kirstein \orcidlink{0000-0002-2549-2115}
}\thanks{These authors contributed equally to this work.}
\affiliation{National High Magnetic Field Laboratory, Los Alamos National Lab, Los Alamos, NM 87545, USA}

\author{Pyeongjae Park \orcidlink{0000-0002-5246-1779}
}\thanks{These authors contributed equally to this work.}
\affiliation{Materials Science and Technology Division, Oak Ridge National Laboratory, Oak Ridge, TN 37831, USA}
\affiliation{Department of Physics and Astronomy, Seoul National University, Seoul 08826, Korea}

\author{Woonghee Cho}
\affiliation{Department of Physics and Astronomy, Seoul National University, Seoul 08826, Korea}

\author{Cristian~D.~Batista \orcidlink{0000-0003-1667-3667}
}\thanks{Correspondence: crooker@lanl.gov, jgpark10@snu.ac.kr, cbatist2@utk.edu}
\affiliation{Department of Physics and Astronomy, University of Tennessee, Knoxville, TN, USA}
\affiliation{Shull Wollan Center—A Joint Institute for Neutron Sciences, Oak Ridge National Laboratory, Oak Ridge, TN, USA}

\author{Je-Geun Park \orcidlink{0000-0002-3930-4226}
}\thanks{Correspondence: crooker@lanl.gov, jgpark10@snu.ac.kr, cbatist2@utk.edu}
\affiliation{Department of Physics and Astronomy, Seoul National University, Seoul 08826, Korea}
\affiliation{Institute of Applied Physics, Seoul National University, Seoul 08826, Korea}

\author{Scott A. Crooker \orcidlink{0000-0001-7553-4718}
}\thanks{Correspondence: crooker@lanl.gov, jgpark10@snu.ac.kr, cbatist2@utk.edu}
\affiliation{National High Magnetic Field Laboratory, Los Alamos National Lab, Los Alamos, NM 87545, USA}

\title{Tunable chiral and nematic states in the triple-$\mathbf{Q}$ antiferromagnet Co$_{1/3}$TaS$_2$}

\begin{abstract}
Complex spin configurations in magnetic materials, ranging from collinear single-\textbf{Q} to non-coplanar multi-\textbf{Q} states, exhibit rich symmetry and chiral properties. However, their detailed characterization is often hindered by the limited spatial resolution of neutron diffraction techniques. Here we employ magnetic circular dichroism (MCD) and magnetic linear dichroism (MLD) to investigate the triangular lattice antiferromagnet Co$_{1/3}$TaS$_2$, revealing three-state ($Z_{3}$) nematicity and also spin chirality across its multi-\textbf{Q} magnetic phases. At intermediate temperatures, the presence of MLD identifies nematicity arising from a single-\textbf{Q} stripe phase, while at high magnetic fields and low temperatures, a phase characterized solely by MCD emerges, signifying a purely chiral non-coplanar triple-\textbf{Q} state. Notably, at low temperatures and small fields, we discover a unique phase where both chirality \textit{and} nematicity coexist. A theoretical analysis based on a continuous multi-\textbf{Q} manifold captures the emergence of these distinct magnetic phases, as a result of the interplay between four-spin interactions and weak magnetic anisotropy. Additionally, MCD and MLD microscopy spatially resolves the chiral and nematic domains. Our findings establish Co$_{1/3}$TaS$_2$ as a rare platform hosting diverse multi-\textbf{Q} states with distinct combinations of spin chirality and nematicity while demonstrating the effectiveness of polarized optical techniques in characterizing complex magnetic textures. 
\end{abstract}

\maketitle

\section{Introduction}

Antiferromagnets, long overshadowed by their ferromagnetic counterparts, are rapidly emerging as a new frontier in condensed matter physics~\cite{jungwirth2016, baltz2018, cheong2020, han2023}. Their complex spin orders offer deep insights into magnetic symmetries, chiralities, and associated emergent phenomena~\cite{smejkal2018, cheong2019, AFM_AHE_2022, rimmler2024}.  Antiferromagnets (AFMs) can exhibit diverse spin configurations, ranging from collinear structures described by a single ordering wave vector $\mathbf{Q}$, to complex non-collinear and non-coplanar multi-$\mathbf{Q}$ spin configurations arising from the superposition of multiple Fourier components with distinct wave vectors $\mathbf{Q}_{\nu}~(\nu=1,2,3..,)$. Notably, multi-$\mathbf{Q}$ states can give rise to topologically non-trivial spin textures~\cite{muhlbauer2009, okubo2012, nagaosa2013, hayami2021} that manifest unique phenomena~\cite{neubauer2009, fert2013, fert2017, jiang2017, kurumaji2019}. Understanding these AFM orders requires precise identification of their magnetic symmetries and potential topological properties, both of which are central themes in modern magnetism research. 

The triangular lattice antiferromagnetism in Co$_{1/3}$TaS$_2$, an intercalated metallic van der Waals system with Co$^{2+}$ spins on 2D triangular lattices (depicted in Fig. 1a), provides an excellent platform for exploring the rich physics associated with competing AFM orders. In this system, conduction-electron-mediated two-spin and four-spin interactions stabilize a non-coplanar triple-$\mathbf{Q}$ ground state below 26.5\,K~\cite{takagi2023, park2023_ncomm}. This state, characterized by ordering wave vectors at the three $M$ points (edges) of the hexagonal Brillouin zone, represents the shortest-wavelength limit of a magnetic skyrmion crystal~\cite{wang2023}. The resulting chiral spin texture can generate a pronounced topological Hall effect (without requiring relativistic spin-orbit coupling), leading to a large Hall conductivity $\sigma_{xy}$ despite vanishing magnetization, and even in the absence of applied magnetic field~\cite{park2023_ncomm, takagi2023}. At intermediate temperatures between 26.5\,K and 38\,K, \CTS{} transitions to a stripe-like single-$\mathbf{Q}$ antiferromagnetic phase~\cite{park2023_ncomm, park2024_INS}, which has been proposed as a promising candidate for realizing discrete three-state ($Z_{3}$) electronic nematicity~\cite{little2020_Fe1/3NbS2, zhang2021_FePS3, Hwangbo2024}, arising from rotational symmetry breaking in a hexagonal lattice. Beyond these zero-field chiral and nematic phases, out-of-plane magnetic fields $H\gtrsim 3.5$\,T induce two additional AFM phases, whose spin configurations and ordering mechanisms remain unknown~\cite{park2022_npj,takagi2023,park2023_ncomm}.

These recent discoveries raise fundamental questions about how magnetic chirality and nematicity evolve across these temperature- and field-induced phases. Prior reports on similar phenomena are scarce, and neutron diffraction methods applied to date struggle to fully distinguish and characterize magnetic chirality and nematicity, due to a lack of spatial resolution and inability to resolve magnetic domains~\cite{park2023_ncomm, takagi2023}. This challenge highlights the need for complementary approaches to investigate the rich ($H,T$) phase diagram of Co$_{1/3}$TaS$_2$, ideally with experimental probes that can illuminate the interplay and potential coexistence of single-$\mathbf{Q}$ and triple-$\mathbf{Q}$ orderings within the same system -- an uncommon feature in real materials. %However, neutron diffraction methods applied to date struggle to characterize both magnetic chirality and nematicity, as its inability to resolve spatial profiles and domain formation leaves ambiguity between single-$\mathbf{Q}$ and triple-$\mathbf{Q}$ scenarios, motivating alternative approaches.

\begin{figure*}
\includegraphics[width=.92\linewidth]{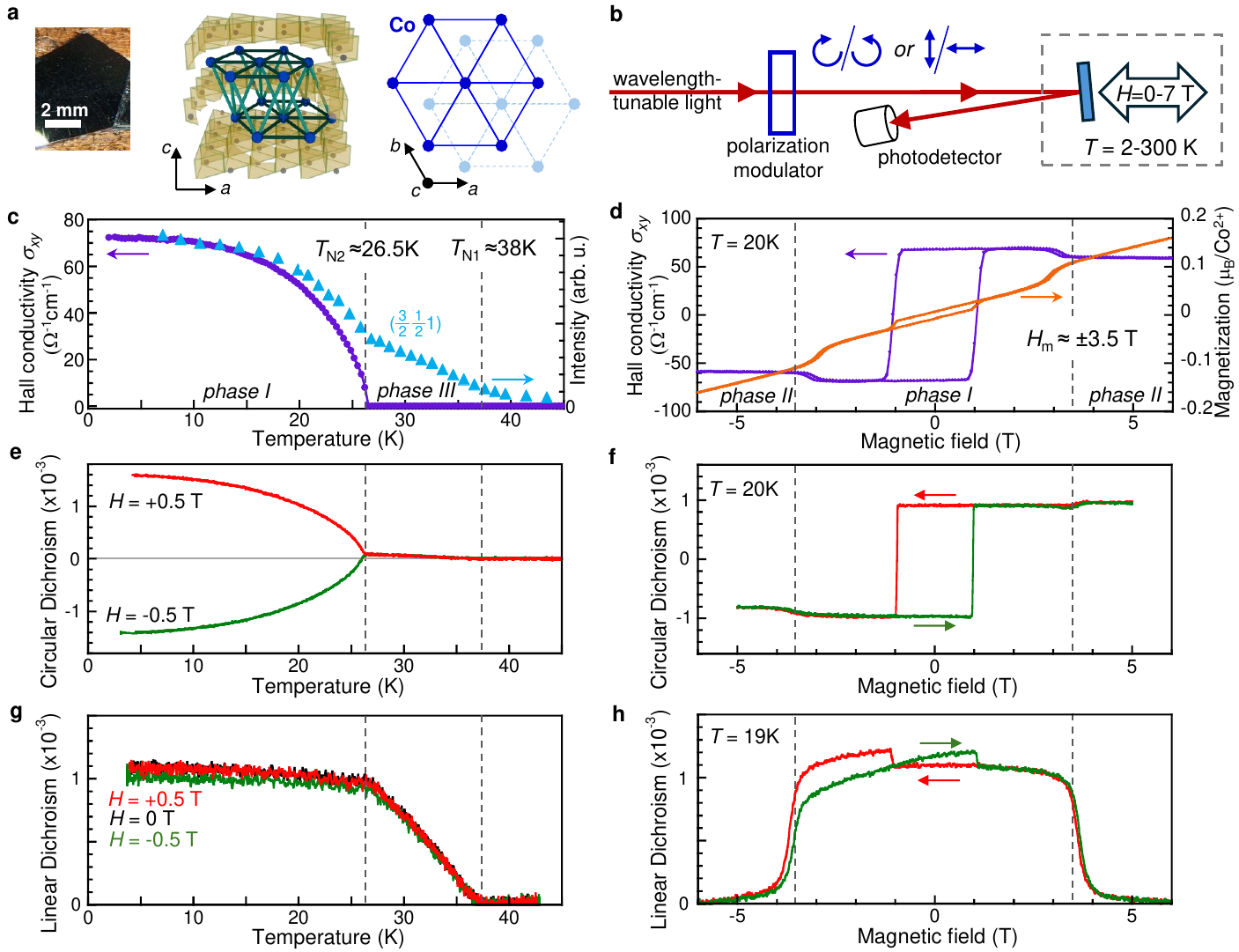}%\\
\caption{\textbf{Optical detection of chiral and nematic antiferromagnetic order in Co$_{1/3}$TaS$_2$.} \textbf{a}, Illustration of the Co$_{1/3}$TaS$_2$ crystal structure. Co ions (blue) intercalate between TaS$_2$ monolayers, forming layers of spins on 2D triangular lattices with ABAB stacking, resulting in a tetrahedral network of Co spins. \textbf{b}, Schematic of the magnetic circular dichroism (MCD) and magnetic linear dichroism (MLD) experiment. Wavelength-tunable light is modulated between right- and left-circular polarization (for MCD) or between linear and cross-linear polarization (for MLD) by a photoelastic modulator, and then reflected from the sample at near-normal incidence and detected by a photodiode. Out-of-plane magnetic fields $H$ to $\pm$7\,T can be applied. MCD measures the normalized reflected intensity difference between right- and left-polarized light, ($I_R - I_L)/(I_R + I_L)$, and is sensitive to, e.g., chiral (triple-$\mathbf{Q}$) AFM order. MLD measures the normalized reflected intensity difference between linear- and cross-linear light, ($I_\phi - I_{\phi + 90^\circ})/(I_\phi + I_{\phi + 90^\circ})$, and is sensitive to, e.g., nematic (single-$\mathbf{Q}$) AFM order with broken in-plane ($C_{2z}$) symmetry. \textbf{c}, The recently-reported Hall conductivity $\sigma_{xy}$ and neutron diffraction intensity from Co$_{1/3}$TaS$_2$ vs. $T$ (from \cite{park2023_ncomm}), showing the onset of single-$\mathbf{Q}$ order below $T_{N1}$=38\,K and triple-$\mathbf{Q}$ (chiral) order below $T_{N2}$=26.5\,K. \textbf{d}, The recently-reported $\sigma_{xy}$ and magnetization $M$ vs. $H$, showing large hysteretic $\sigma_{xy}$ despite small $M$, and additional metamagnetic transitions at $H_m \approx \pm 3.5$\,T whose nature is not known to date. \textbf{e,f}, MCD studies vs. $T$ and $H$ confirm the emergence of chiral AFM order below $T_{N2}$, and a large $H$-dependent hysteresis, closely following the electrical transport measurements of $\sigma_{xy}$ shown in the panels above. \textbf{g,h}, MLD studies vs. $T$ and $H$ reveal the emergence of \textit{nematic} order below $T_{N1}$ and its saturation below $T_{N2}$, and reveal that this nematicity exists only at low fields $|H|<H_m$. (Small linear backgrounds have been removed from $H$-dependent data of $\sigma_{xy}$, MCD, and MLD). These data point to nematic (single-$\mathbf{Q}$) AFM order at high $T$, purely chiral (``equilateral triple-$\mathbf{Q}$'') AFM order at low $T$ and high $H$, and to the coexistence of both chiral \textit{and} nematic (``non-equilateral triple-$\mathbf{Q}$'') AFM order at low $T$ and low $H$.  All optical data acquired using 650~nm light.}
\label{fig:intro}
\end{figure*}

Here we demonstrate that optical methods for magnetic circular dichroism (MCD) and magnetic linear dichroism (MLD) directly provide spatially-resolved measurements of chirality and nematic symmetries, respectively, of the different antiferromagnetic orders in Co$_{1/3}$TaS$_2$. These techniques reveal nematicity arising from single-$\mathbf{Q}$ order at intermediate $T$, purely chiral triple-$\mathbf{Q}$ order at low $T$ and large $H$, and -- crucially -- the coexistence of \textit{both} chiral and nematic (``non-equilateral triple-$\mathbf{Q}$'') order at low $T$ and $H$.  Our findings establish Co$_{1/3}$TaS$_2$ as a rare material hosting diverse multi-\textbf{Q} states, and highlight the power of magneto-optical techniques for characterizing complex magnetic textures. %within a framework for continuous multi-\textbf{Q} magnetic ordering, accurately predicting the ($H,T$) phase diagram and the spin configurations in each phase. Finally, we employ MCD and MLD microscopy to directly image the chiral and $Z_{3}$ nematic domains and unravel their superimposed polarization directors. As a rare realization of a continuous multi-\textbf{Q} manifold, our results demonstrate that magneto-optical approaches offer viable routes for exploring multi-\textbf{Q} antiferromagnetism with interrelated chiral and nematic properties.

\section{Results} 

\subsection{Antiferromagnetic phases in \CTS{}}
\CTS{} consists of 2D triangular lattices of Co spins, intercalated within the van der Waals gaps of 2$H$-TaS$_2$ (see Fig.~\ref{fig:intro}a). Nearest-neighbour interlayer couplings and AB stacking create an effective tetrahedral network of Co, which orders antiferromagnetically. Figures~\ref{fig:intro}c,d briefly review its diverse AFM phases, revealed by recent transport, neutron scattering, and magnetization studies~\cite{park2022_npj, takagi2023, park2023_ncomm}. When cooled in $H=0$, AFM order first appears below $T_{\mathrm{N1}}$\,=\,38\,K, where neutron studies identify collinear single-\textbf{Q} (stripe) order~\cite{park2024_INS} with Co spins aligned out-of-plane \cite{park2023_ncomm, takagi2023}. Upon further cooling, a new AFM ground state emerges below $T_{\mathrm{N2}}$\,=\,26.5\,K that is characterized by a large spontaneous Hall conductivity $\sigma_{xy}$, despite vanishingly small zero-field magnetization ($M_z$\,=\,0.01$\mu_B$/Co$^{2+}$). This low-temperature phase is identified as a chiral non-coplanar ``tetrahedral'' triple-\textbf{Q} order. Electrons moving within this chiral spin texture accumulate a geometric (Berry) phase that generates an emergent magnetic field, resulting in substantial $\sigma_{xy}$ despite tiny $M_z$ -- generally referred to as a topological Hall effect~\cite{ye1999, taguchi2001, bruno2004, neubauer2009}. These triple-$\mathbf{Q}$ and single-$\mathbf{Q}$ phases are designated Phases I and III, respectively ~\cite{takagi2023}. 

Crucially, applying out-of-plane $H$ larger than $H_m \approx3.5$\,T) induces additional metamagnetic transitions, observed as small jumps in $\sigma_{xy}$ and $M_{z}$ (see Fig.~\ref{fig:intro}d) \cite{park2022_npj, takagi2023}. These transitions lead to two additional phases—Phase II and Phase IV—presumably with distinct spin configurations and ordering mechanisms that remain unresolved.  %These field-induced phases, denoted II and IV, presumably involve spin reconfigurations to distinct AFM ground states. However, their order parameters and nature of the metamagnetic transition remain unresolved. 

To directly probe the magnetic order of these phases we employ MCD and MLD -- optical techniques that selectively measure chiral and nematic properties, respectively. The experiment is depicted in Fig.~\ref{fig:intro}b (see caption for details, and Methods).  Traditionally, MCD has been used to detect ferromagnetic order, as it probes the nonzero off-diagonal optical conductivity $\sigma_{xy}(\omega)$, which also underpins the dc Hall effect at $\omega$$\approx$0. However, recent discoveries of antiferromagnets with non-trivial spin orders whose (reduced) symmetry allows for $\sigma_{xy}(\omega)$ -- such as non-collinear Mn$_3$Sn~\cite{kubler2014, chen2014, Nakatsuji2015, suzuki2017} or here for non-coplanar Co$_{1/3}$TaS$_2$ -- suggests that MCD can directly probe such AFM order \cite{higo2018, balk2019, feng2020, kato2023, li2024}. 

Complementing MCD, MLD has recently been demonstrated as an effective probe of stripe-like single-\textbf{Q} AFM order in 2D hexagonal magnets \cite{zhang2021_FePS3, little2020_Fe1/3NbS2, Hwangbo2024}. Such order breaks the rotational symmetry of the underlying crystal, inducing in-plane ($C_{2z}$) anisotropy of the optical conductivity and giving rise to three-state $Z_3$ nematic order. Together, MLD and MCD provide a powerful framework for characterizing multi-$\mathbf{Q}$ magnetic states in \CTS{}.

\subsection{Detecting chiral and nematic AFM order with MCD and MLD}
To establish MCD as a probe of the chiral triple-\textbf{Q} order, we measure its temperature and field dependence. Fig.~\ref{fig:intro}e shows that strong MCD signals emerge below \TNt{}, closely following the behavior of $\sigma_{xy} (T)$ from transport measurements (Fig. 1c).  Furthermore, Fig.~\ref{fig:intro}f shows that MCD exhibits a marked hysteresis with $H$, in agreement with prior transport studies shown in Fig. 1d~\cite{park2022_npj, park2023_ncomm, takagi2023}. These findings confirm that MCD is sensitive to chiral AFM order in \CTS{}. The scalar spin chirality of this triple-$\mathbf{Q}$ state, defined as $\chi_{ijk} = S_i \cdot (S_j \times S_k)$ where $i,j,k$ label spins around any triangular plaquette in the lattice, can be switched between time-reversed configurations by $H$ ($\approx\pm$1\,T at 20\,K), giving opposite $\sigma_{xy}(\omega)$ and MCD. The sharp switching behavior indicates high sample quality and uniform reversal of the chiral order. Additionally, MCD signals reveal the meta-magnetic transition at $H_m \approx \pm 3.5$\,T, in close correspondence with transport and magnetization measurements (Fig. 1d; see also Supplementary Figs. S1 and S2). The correspondence of MCD and Hall conductivity suggests their common origin: the real-space Berry curvature generated by chiral triple-$\mathbf{Q}$ magnetism~\cite{park2023_ncomm, takagi2023}. 

In marked contrast, Fig.~\ref{fig:intro}g shows that MLD emerges below $T_{\mathrm{N1}}$\,=\,38\,K, coinciding with the onset of stripe-like single-\textbf{Q} AFM order. As shown in recent studies of 2D hexagonal antiferromagnets FePS$_3$ and Fe$_{1/3}$NbS$_2$~\cite{little2020_Fe1/3NbS2, zhang2021_FePS3, Hwangbo2024}, collinear single-$\mathbf{Q}$ stripe order can generate substantial MLD due to rotational symmetry breaking and consequent asymmetry of the in-plane optical conductivity. The MLD in \CTS{} saturates at a large value below \TNt{}, indicating the persistence of nematicity in the low-temperature phase. The MLD does not vary significantly when cooled in small $\pm H$, or in $H=0$. Importantly, the simultaneous presence of both MLD and MCD at $T< T_{\mathrm{N2}}$ (Figs. 1e,g) provides direct evidence that Phase I exhibits both chiral \textit{and} nematic order.  

Field-dependent MLD further clarifies the nature of the puzzling metamagnetic phase transition at $H_m \approx \pm$3.5\,T. As shown in Fig.~\ref{fig:intro}h, the MLD is large at low $H$ but vanishes when $|H|>H_m$, indicating the \textit{disappearance} of nematicity.  Meanwhile, the MCD remains large (see Fig. 1f), demonstrating that the $H_m$ separates a nematic-chiral phase (Phase I) from a purely chiral phase (Phase II). Notably, small jumps in the MLD signal coincide with chirality reversals, likely arising in part from cross-talk between linear and circular dichroism signals. This cross-talk also generates the very small MCD signal between $T_{\mathrm{N1}}$ and $T_{\mathrm{N2}}$ in Fig. 1e (see Supplementary Fig. S2). 

\begin{figure*}[t]
\includegraphics[width=.93\linewidth]{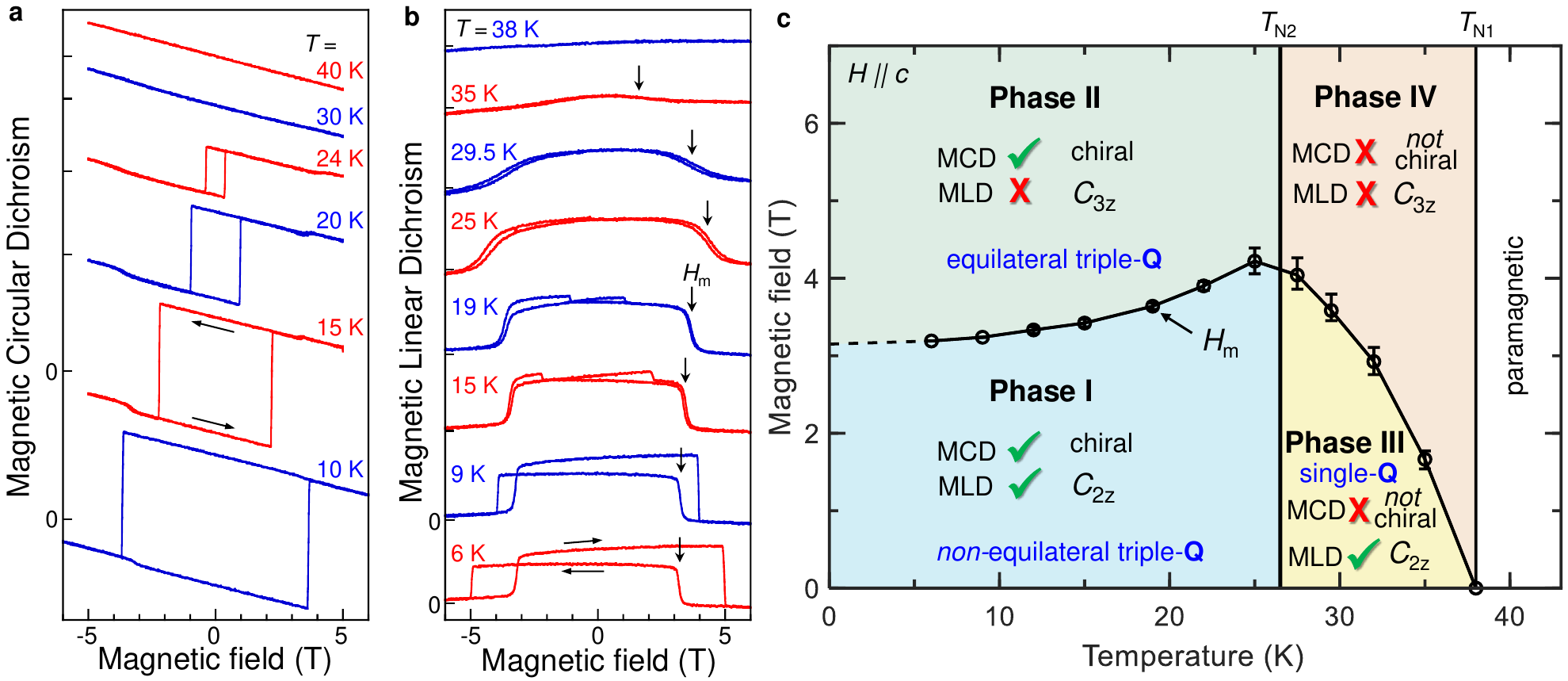}%\\
\caption{\textbf{Mapping out the antiferromagnetic phase diagram of Co$_{1/3}$TaS$_2$.} \textbf{a}--\textbf{b}, MCD and MLD versus $H$, at different temperatures, showing how signatures of chiral and nematic AFM order, revealed by MCD and MLD respectively, emerge and/or vanish with changing $T$ and $H$ (curves offset for clarity). All data acquired using $\lambda$=650~nm. \textbf{c}, Magnetic phase diagram of Co$_{1/3}$TaS$_2$ based on these optical data, where the four different Phases I-IV are defined by the presence or absence of anomalous MCD (chiral AFM order), and MLD (nematic AFM order).}  
\label{fig:hysttemperature}
\end{figure*} 

\subsection{Mapping the magnetic phase diagram}

Figures~\ref{fig:hysttemperature}a,b show field-dependent MCD and MLD at different temperatures. As previously observed in $\sigma_{xy}$~\cite{park2022_npj, takagi2023}, both the chiral switching field $H_{\textrm{c}}$ and the amplitude of the hysteresis loops increase rapidly below \TNt{}. Importantly, the disappearance of MLD at large $H$ persists for all $T<T_{\mathrm{N1}}$, reinforcing the phase distinction [see also MLD($T$) scans at fixed $H$ in Supplementary Fig. S3]. Using these magneto-optical measurements, we construct the ($H,T$) phase diagram in Fig.~\ref{fig:hysttemperature}c, defining Phases I-IV based on the presence or absence of MCD (chirality) and MLD (nematicity). 

The MCD results confirm that Phases I and II exhibit chiral triple-\textbf{Q} order, while the high temperature Phases III and IV do not. More importantly, MLD offers new insights into the symmetries of Phases I, II, and IV. The absence of rotational $C_{3z}$ symmetry in Phase I, revealed by MLD, contradicts the recently-proposed $C_{3z}$-symmetric triple-\textbf{Q} ground state~\cite{takagi2023}. Instead, the presence of both MCD and MLD in Phase I suggests a triple-\textbf{Q} state with \textit{broken} $C_{3z}$ rotational symmetry -- that is, with both chiral \textit{and} nematic AFM order. As described in the next section, such novel states can be realized through intermediate spin configurations involving coexisting single-\textbf{Q} and triple-\textbf{Q} components~\cite{haldar2021}. Finally, we emphasize that Phase II, which exhibits only MCD, appears to realize the $C_{3z}$-symmetric triple-$\mathbf{Q}$ state that was originally proposed for Phase I.  % revealed by MLD at low fields, and its restoration for $H>H_m$ offer new insights into the metamagnetic transition and the symmetries of Phases I, II, and IV (phase III is already known to host single-$\mathbf{Q}$ stripe order \cite{park2024_INS}). The recently-proposed $C_{3z}$-symmetric triple-$\mathbf{Q}$ ground state for Phase I \cite{takagi2023} contradicts with our observation of large MLD signal (Fig.~\ref{fig:intro}g), whereas Phase II (with vanishing MLD) appears more consistent with such a state. Given that Phase I’s triple-$\mathbf{Q}$ nature is firmly established by nonzero MCD and previous neutron and transport studies \cite{park2024_INS, park2023_ncomm, takagi2023} (Fig.~\ref{fig:hysttemperature}c), our findings point to a triple-$\mathbf{Q}$ order with \textit{broken} (i.e., $C_{2z}$) rotational symmetry as the only viable scenario for antiferromagnetic ground state in Phase I. 

\begin{figure*}
\includegraphics[width=.98\linewidth]{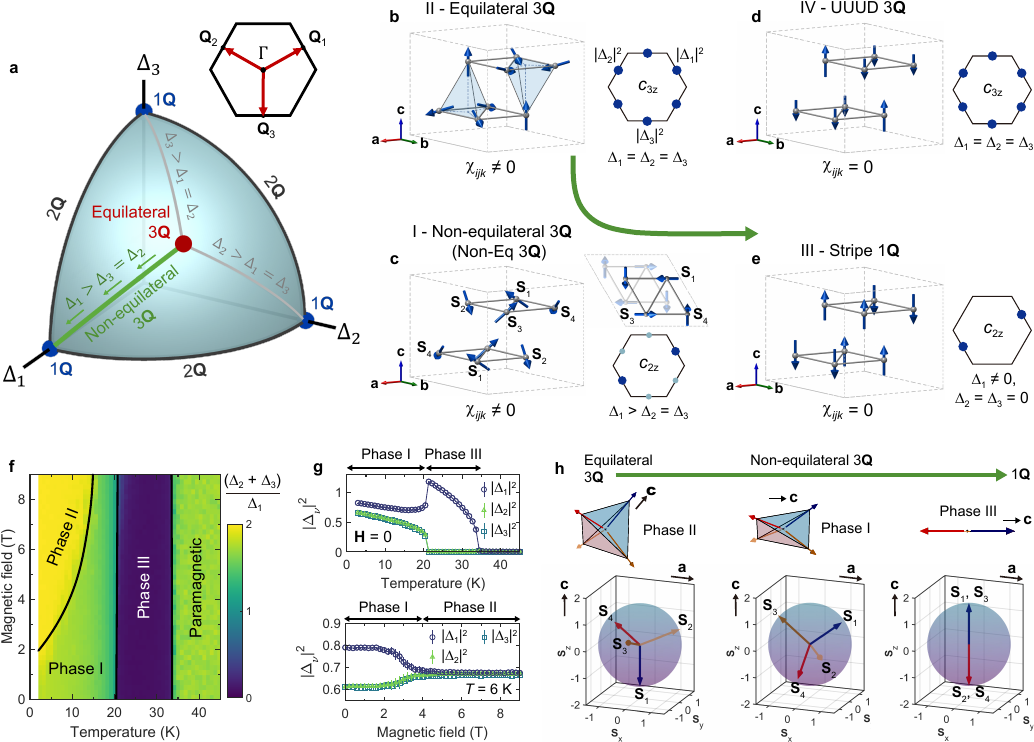}%\\
\caption{\textbf{Theoretical model of the $H,T$ phase diagram based on continuous multi-Q manifolds.} \textbf{a}, Continuous manifold of multi-$\mathbf{Q}$ $M$-orderings, depicted as a variational space on a spherical shell. \textbf{b}--\textbf{e}, Magnetic structures consistent with the symmetry and topological properties suggested by the combination of MCD and MLD measurements. The size and colour of the blue dots on the right side of each panel represent the magnitude of the Fourier components ($|\Delta_{\nu}|^2$). \textbf{f}--\textbf{g}, Theoretical temperature-field ($\mathbf{H} \parallel \mathbf{c}$) phase diagram derived from classical Monte-Carlo simulations of the realistic anisotropic spin model for \CTS{} (Eq.~\eqref{eq:Hamiltonian} and ~\eqref{eq:H_aniso_main}). The colour code in \textbf{f} indicates $r_{\Delta} \equiv \frac{\Delta_{2} + \Delta_{3}}{\Delta_{1}}$, where $\Delta_{3} \leq  \Delta_{2} \leq \Delta_{1}$. Spin configurations of the Phases I--III are depicted in \textbf{c}, \textbf{b}, and \textbf{e}, respectively. \textbf{h}, Evolution of the four spin-sublattice configurations throughout Phases II, I, and III, transitioning from an equilateral triple-$\mathbf{Q}$ (3\textbf{Q}) state to a stripe single-$\mathbf{Q}$ (1\textbf{Q}) state via a non-equilateral tetrahedral alignment.}
\label{fig:theory}
\end{figure*}

\subsection{Continuous multi-\textbf{Q} manifold model}

These observations motivate a continuous multi-\textbf{Q} manifold model that smoothly interpolates between single-\textbf{Q} and $C_{3z}$-symmetric triple-\textbf{Q} states, and which naturally captures the evolution of spin chirality and nematicity in \CTS{}. For $M$-ordering wave vectors, this continuous interpolation can be modelled by the following manifold (see Fig.~\ref{fig:theory}a):

\begin{equation}
{\mathbf{S}}({\mathbf{r}}) = \sum_{\nu=1}^{3} \tilde{\mathbf{S}}_{{\mathbf{Q}}_{\nu}} \cos({{\mathbf{Q}}_{\nu} \cdot {\mathbf{r}}}),
\label{eq:manifold_main}
\end{equation}
where ${\mathbf{r}}$ is a lattice vector on the 2D triangular lattice, and ${\mathbf{S}}({\mathbf{r}})$ denotes a classical spin vector. The derivation of Eq.~\eqref{eq:manifold_main} and its validity for \CTS{} is described in Supplementary  Section II A. By adjusting the relative magnitudes of the three Fourier components ($|\tilde{\mathbf{S}}_{{\mathbf{Q}}_{\nu}}|\equiv \Delta_{\nu}$) while maintaining their orthogonality and a constant total magnitude $\Delta_{1}^2+\Delta^2_{2}+\Delta^2_{3}$, this ansatz spans a spherical surface in ($\Delta_{1},\Delta_{2},\Delta_{3}$) phase space (see Fig.~\ref{fig:theory}a), which interpolates between all possible single-, double-, and triple-$\mathbf{Q}$ spin configurations with fixed $|{\mathbf{S}}({\mathbf{r}})|$. However, Phase IV is not captured by Eq.~\eqref{eq:manifold_main} and will be discussed separately.

The case $\Delta_1 = \Delta_2 = \Delta_3$ (red circle in Fig.~\ref{fig:theory}a) yields a chiral and $C_{3z}$-symmetric ground state, corresponding to Phase II (Fig.~\ref{fig:theory}b), where the four spin sublattices align along the principal axes of an equilateral tetrahedron -- referred to as ``equilateral triple-$\mathbf{Q}$''. 
In contrast, a single non-zero $\Delta_{\nu}$ (blue circles in Fig.~\ref{fig:theory}a) corresponds to the single-$\mathbf{Q}$ stripe order of Phase III (Fig.~\ref{fig:theory}e). Intermediate states arise when $0 < \Delta_i = \Delta_j < \Delta_k$ for indices $i,j,k \in {1,2,3}$ (green lines throughout Fig.~\ref{fig:theory}). The broken $C_{3z}$ symmetry is evident from the unequal $\Delta_{\nu}$ magnitudes (Fig.~\ref{fig:theory}c), resulting in the four spin sublattices spanning a \textit{distorted} tetrahedron, which we term ``non-equilateral triple-$\mathbf{Q}$'' (or Non-Eq 3\textbf{Q}). From this perspective, thermal fluctuations and $H$ cause the magnetic ground state to evolve smoothly within this multi-$\mathbf{Q}$ manifold, effectively controlling both the spin chirality and $Z_{3}$ nematicity in \CTS{}.

While the emergence of a continuous multi-$\mathbf{Q}$ manifold is rarely observed in real systems, our theoretical analysis adds strong evidence for its feasibility based on a realistic spin Hamiltonian for \CTS{}. The isotropic low-energy Hamiltonian introduced in previous studies~\cite{park2023_ncomm, park2024_INS}, which incorporates bilinear Heisenberg and four-spin interactions, fails to fully capture the field- and temperature-driven multi-$\mathbf{Q}$ manifold. While a more generalized model could include additional four-spin terms with varying forms~\cite{sharma2023}, we find that a simple real-space biquadratic term ($\mathcal{\hat H}_{\rm bq}$) captures the most general classical ground state of the multi-$\mathbf{Q}$ $M$-ordering and its long-wavelength fluctuations:

\begin{eqnarray}
{\hat{\cal H}_{\mathrm{bq}}} 
  &=& {K}\sum_{{\mathbf{r}}, {\bm{\delta}}_1} (\hat {\mathbf{S}}_{\mathbf{r}} \cdot \hat {\mathbf{S}}_{{\mathbf{r}}+{\bm{\delta}}_1})^{2},
\label{eq:Hamiltonian}
\end{eqnarray}
where $\bm{\delta}_{1}$ is the vector connecting nearest-neighbour sites (derived in Supplementary Section II.B--C). Notably, ${\hat{\cal H}_{\mathrm{bq}}}$ with $K>0$ successfully captures the single-$\mathbf{Q}$ to triple-$\mathbf{Q}$ transition at \TNt{}~\cite{park2023_ncomm, park2024_INS}. However, it incorrectly predicts a $C_{3z}$-symmetric triple-$\mathbf{Q}$ ground state at $T=H=0$ (Phase I), and fails to explain any field-induced metamagnetic transition. 

Magnetic anisotropy in \CTS{}, though smaller than $K$ in magnitude, is suggested by the finite magnon energy gap in Phase I and by the out-of-plane spin configuration in Phase III~\cite{park2023_ncomm, park2024_INS} (see Supplementary Section II.D). To account for this, we include single-ion easy-axis anisotropy (${\hat{\cal H}}_{\rm SI}$) and bond-dependent exchange anisotropy (${\hat{\cal H}}_{\rm \pm\pm}$) in the isotropic model:

\begin{eqnarray}
{\hat{\cal H}_{\mathrm{SI}}} &=& A \sum_{\mathbf{r}}  (\hat{S}^{z}_{\mathbf{r}})^2, \nonumber \\
{\hat{\cal H}_{\pm\pm}} &=& \sum_{{\langle i,j \rangle}_{1}} 2J_{\pm\pm} \Big[ \left( S_i^x S_j^x - S_i^y S_j^y \right) \cos \phi_\alpha \nonumber \\ 
&-& \left( S_i^x S_j^y + S_i^y S_j^x \right) \sin \phi_\alpha \Big],
\label{eq:H_aniso_main}
\end{eqnarray}
where ${{\langle i,j \rangle}_{1}}$ runs over nearest-neighbour bonds, $x \parallel$  $a$-axis, and $\phi_{\alpha} \in \{0, 2\pi/3, 4\pi/3\}$ is the angle between a bond vector ($i \to j$) and the $a$-axis. The resultant phase diagram at $T=H=0$ spanned by $J_{\pm\pm}$ and $A$ is shown in Fig.~\ref{Sfig:multiQpdg}b and ~\ref{Sfig:3Dpdg}, where we effectively visualize it through a quantity $r_{\Delta} \equiv \frac{\Delta_{2} + \Delta_{3}}{\Delta_{1}}$, where $\Delta_{3} \leq  \Delta_{2} \leq \Delta_{1}$. Introducing a non-zero $A$ immediately changes the ground state from equilateral triple-$\mathbf{Q}$ ($r_{\Delta}=2$, Fig.~\ref{fig:theory}b) to non-equilateral triple-$\mathbf{Q}$ ($1<r_{\Delta}<2$, Fig.~\ref{fig:theory}c), consistent with the observed MLD in Phase I. Thus, an intermediate triple-$\mathbf{Q}$ ground state on the continuous manifold is a natural explanation for Phase I under the presence of magnetic anisotropy.

Further exploration of temperature- and field-dependent ground states was performed using classical Monte Carlo simulations (see Methods, Supplementary Note II.D). The results in Fig.~\ref{fig:theory}f--h, presented as $\Delta_{\nu}$ and $r_{\Delta}$, reveal: i) a non-equilateral triple-$\mathbf{Q}$ ground state ($1<r_{\Delta}<2$) in Phase I, ii) a field-induced equilateral triple-$\mathbf{Q}$ state ($r_{\Delta}=2$) in \Tb{} (Phase II), and iii) a single-$\mathbf{Q}$ state ($r_{\Delta}=0$) in \Ti{} (Phase III). The consistency with Fig.~\ref{fig:hysttemperature} strongly supports the continuous multi-$\mathbf{Q}$ manifold interpretation, linking single-$\mathbf{Q}$ and $C_{3z}$-symmetric triple-$\mathbf{Q}$ orderings with distinct chiral and nematic properties. Ground state visualizations are shown in Fig.~\ref{fig:theory}h.

We now discuss Phase IV and the limitations of our model. The absence of both MLD and MCD in Phase IV implies preserved $C_{3z}$ symmetry, suggesting a non-chiral triple-$\mathbf{Q}$ state. A likely configuration is the up-up-up-down (UUUD) structure, a collinear triple-$\mathbf{Q}$ ordering (Fig.~\ref{fig:theory}d) observed in triangular lattice antiferromagnets under a magnetic field~\cite{haley2020}. In  classical simulations, the UUUD phase requires a magnetization of half the saturation value ($M=1.5\mu_{\mathrm{B}}$/Co$^{2+}$), thereby appearing at higher fields in our model compared to $H_{m} \sim$3.5\,T (Phase IV exhibits $M=0.2\mu_{\mathrm{B}}$/Co$^{2+}$ at 7\,T). We attribute this discrepancy to longitudinal spin fluctuations not captured in our classical model. Indeed, the ordered magnetic moment of $1.3\mu_{\mathrm{B}}$/Co$^{2+}$ observed at 3\,K--much smaller than the classical $3\mu_{\mathrm{B}}$/Co$^{2+}$--indicates substantial spin fluctuations~\cite{park2023_ncomm}. A possible consequence of strong fluctuations in triangular lattice antiferromagnets is a site-dependent renormalization of ordered moments~\cite{zhu2024}, e.g., the three ``up'' moments could become shorter than the single ``down'' moment (Fig.~\ref{fig:theory}e), reducing net magnetization. Precise determination of Phase IV requires future neutron diffraction measurements under magnetic field.

\subsection{Imaging nematic and chiral antiferromagnetic domains}
\begin{figure*}
\includegraphics[width=.92\linewidth]{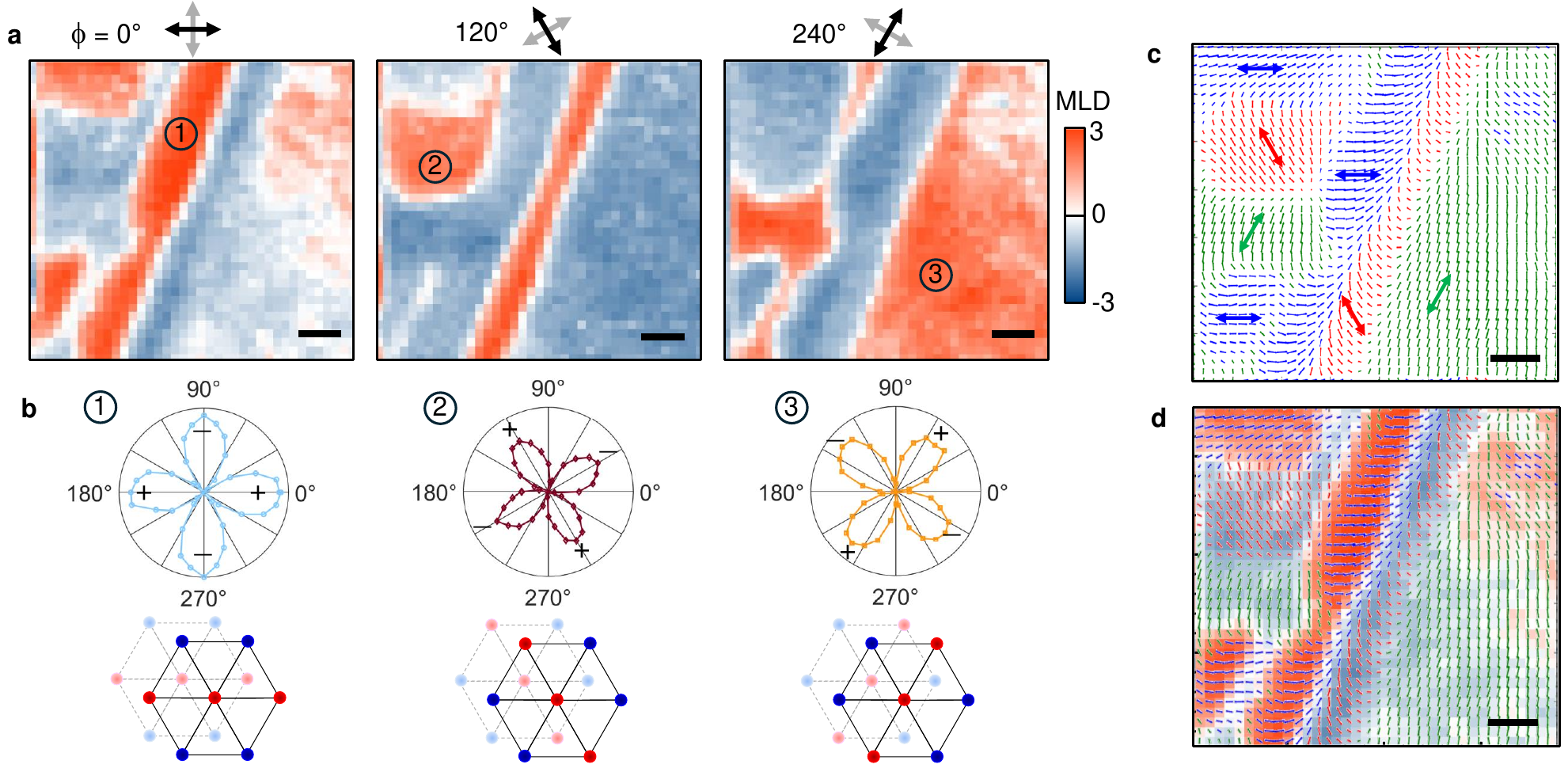}%\\
\caption{\textbf{Spatially-resolved images of nematic (single-$\mathbf{Q}$) AFM domains in Co$_{1/3}$TaS$_2$}, acquired via scanning MLD microscopy. \textbf{a}, Three $76 \times 70\,\mu \rm{m}$ images of MLD at $T\approx 27$\,K and $H$=0 and (in phase III), using linearly-polarized light modulated between $\phi =0^{\circ}/90^{\circ}$, $120^{\circ}/210^{\circ}$, and $240^{\circ}/330^{\circ}$ with respect to the Co$_{1/3}$TaS$_2$ $a$-axis. Depending on $\phi$, different spatial regions show large and positive MLD signal. (Note: from each image, an MLD image in the nonmagnetic phase at 38\,K was subtracted, reducing background signals). All scale bars are 10$\mu$m. \textbf{b}, Polar plots of the measured MLD versus $\phi$, measured at the three locations indicated. These polar plots demonstrate $C_{2z}$ symmetry of the single-$\mathbf{Q}$ (stripe) AFM order, mutually rotated by 120$^\circ$, and depicted in the diagrams below where red/blue dots indicate Co spins oriented into/out of the page. \textbf{c}, Spatial map of the measured nematic director, revealing domains oriented along the indicated directions. \textbf{d}, Nematic director map overlaid with the first MLD image shown in panel \textbf{a}. }
\label{fig:mld-maps}
\end{figure*}

A notable benefit of optical methods is the ability to directly image real-space profiles of different magnetic states and their domains. In Figs. 4 and 5, we employ MLD and MCD microscopy (see Methods) to spatially resolve, respectively, the discrete $Z_{3}$ nematic domains arising from single-$\mathbf{Q}$ AFM order, and the binary domains associated with positive and negative spin chirality.  Figure~\ref{fig:mld-maps}a shows three MLD images of the same area, acquired in Phase III (single-$\mathbf{Q}$), using probe light linearly polarized at $\phi$= 0$^{\circ}$, 120$^{\circ}$, and 240$^{\circ}$ with respect to the crystal $a$-axis. The images reveal distinct regions with strong positive MLD (red) and weaker negative MLD (light blue). Importantly, the red regions are \textit{different} in each image. Largest MLD occurs when $\phi$ aligns parallel to the stripes of single-$\mathbf{Q}$ magnetic order. The images therefore indicate three different single-$\mathbf{Q}$ domains where the stripes align along 0$^{\circ}$, 120$^{\circ}$, and 240$^{\circ}$, as expected from a triangular lattice. The $C_{2z}$ symmetry of the domains is demonstrated in Fig.~\ref{fig:mld-maps}b, where MLD at the three indicated locations is measured versus $\phi$. Each polar plot exhibits $C_{2z}$ in-plane anisotropy, oriented approximately along 0$^{\circ}$, 120$^{\circ}$, and 240$^{\circ}$ (see also Fig. S4). These images confirm $Z_3$ nematic order in \CTS{}, induced by single-$\mathbf{Q}$ AFM order, with different nematic domains having ordering wave vectors $\mathbf{Q}_{\nu}$ related by $\pm 120^{\circ}$ rotation.

Analyzing the MLD at each position yields a spatial map of the nematic director (Fig.~\ref{fig:mld-maps}c). The directors indicate the angle $\phi$ giving maximum MLD signal, and have lengths proportional to the MLD magnitude. For clarity, blue, red, and green directors represent orientations closest to 0$^{\circ}$, 120$^{\circ}$, and 240$^{\circ}$. This classification scheme remains consistent at different locations on the sample (see Fig. S5). Notably, the nematic domains can be quite large, extending nearly 1\,mm.  

Interestingly, repeated thermal cycling into the paramagnetic state, even staying at 300~K for days or weeks, did not affect the nematic domain patterns. It suggests that they are pinned by extrinsic factors that locally break the underlying hexagonal crystal symmetry, such as local strain. Similarly, Fe$_{1/3}$NbS$_2$ and FePSe$_{3}$ have also shown biasing of an underlying three-state AFM nematicity by uniaxial strain \cite{little2020_Fe1/3NbS2,Hwangbo2024}. Furthermore, the MLD signals remain essentially unchanged below \TNt{} where spin chirality also emerges (\textit{cf.} Fig. 1g), suggesting a smooth transition between the nematic order in Phase III and Phase I. This implies that at \TNt{}, single-$\mathbf{Q}$ domains with $\Delta_i\neq0$ and $\Delta_{j,k}=0$ ($i,j,k\in \{1,2,3\}$) transform continuously into  non-equilateral triple-$\mathbf{Q}$ domains with $\Delta_i>\Delta_{j}=\Delta_{k}>0$.

\begin{figure}
\includegraphics[width=.99\linewidth]{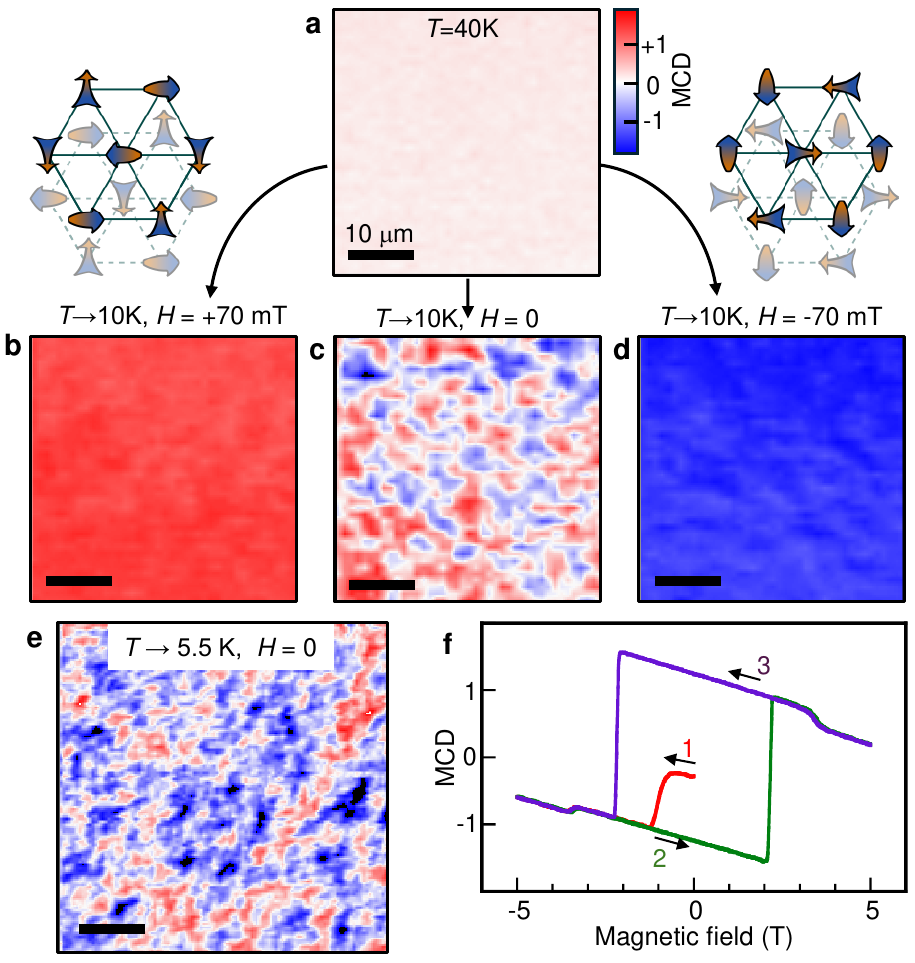}%\\
\caption{\textbf{Spatially resolving spontaneously-forming chiral AFM domains in Co$_{1/3}$TaS$_2$, via MCD microscopy}. \textbf{a}, $40 \times 40\,\mu \rm{m}$ MCD image at $T=40$\,K (in the paramagnetic state) shows no signal. \textbf{b-d}, Same, but after cooling slowly to $T=10$\,K (Phase I) in $H$=+70\,mT, 0\,mT, and -70\,mT. While cooling below $T_{N2}$ in small $H$ readily poles the sample to uniform positive or negative chirality, cooling in zero field results in a dense pattern of spontaneously-formed chiral domains.  Scale bars are 10\,$\mu$m. The diagrams depict two non-equilateral triple-$\mathbf{Q}$ spin configurations related by time-reversal, which have opposite scalar spin chirality but the same nematicity.  Small/large arrowheads are canted into/out of the plane. \textbf{e}, A $50 \times 50\,\mu \rm{m}$ MCD image taken at a different location on the sample, following a more rapid zero-field cooldown to 5.5\,K. \textbf{f}, MCD($H$) scan after initially cooling the sample to 15\,K in $H \approx 0$ (i.e., starting from a configuration with chiral domains). The probe beam is large (1\,mm).  $H$ is ramped from $0 \rightarrow -5\rightarrow +5 \rightarrow -5$\,T. The initial magnetization curve saturates quickly, at a field much smaller than the chiral switching field ($\approx$2\,T) of the main hysteresis loop.}
\label{fig:mcd-maps}
\end{figure}  

Finally, we use MCD microscopy to study spontaneous formation of chiral AFM domains (Fig.~\ref{fig:mcd-maps}). MCD images in the paramagnetic state (40\,K, Fig.~\ref{fig:mcd-maps}a) show no signal, as expected.  However, Figs.~\ref{fig:mcd-maps}b and \ref{fig:mcd-maps}d show that Co$_{1/3}$TaS$_2$ can be completely poled to a positive or negative chiral state by cooling into Phase I in applied $H$ as small as $\pm 70$\,mT. Most importantly, Fig.~\ref{fig:mcd-maps}c shows that small and irregularly-shaped chiral domains spontaneously form when cooled in $H$=0. Their characteristic size is much smaller than the nematic domains. Also in marked contrast to the nematic domains, the chiral domain pattern changes randomly after thermally cycling above \TNo{}, suggesting that the underlying nematicity of the non-equilateral triple-$\mathbf{Q}$ order does not bias the handedness of spontaneously-forming chiral domains. Figure~\ref{fig:mcd-maps}e shows chiral domains at a different location, again showing a random pattern. These images confirm that Phase I cannot be a phase-separated mixture of coexisting single-\textbf{Q} and triple-\textbf{Q} ground states. Finally, Fig.~\ref{fig:mcd-maps}f shows that, starting from a random configuration, the chiral domains are readily poled by fields much less than the coercive field required to switch a fully polarized sample, suggesting weak pinning. 

\section{Summary}
By combining magneto-optical techniques with theoretical analysis, we have elucidated the coexisting chiral and nematic properties in the multi-$\mathbf{Q}$ antiferromagnet \CTS{}. Through temperature- and field-dependent MCD and MLD measurements, we identified the four distinct antiferromagnetic phases, each defined by the presence or absence of spin chirality and nematicity. Thermal fluctuations suppress chirality above \TNt{}, favouring collinear magnetic order, while magnetic fields restore the $C_{3z}$ rotational symmetry and thus suppressing nematicity. These observations are captured by a minimal spin model incorporating four-spin interactions and magnetic anisotropy, which gives rise to a continuous multi-$\mathbf{Q}$ manifold. Real-space imaging using MCD and MLD microscopy revealed robustly-pinned nematic domains with $Z_3$ nematic directors, alongside much smaller, irregularly shaped chiral domains that are easily reoriented by modest magnetic fields—suggesting a weaker pinning mechanism. These contrasting domain behaviors highlight the distinct natures of nematic and chiral order in \CTS{}. Overall, this work demonstrates the efficacy of magneto-optical techniques for characterizing symmetry and chirality in complex spin textures, suggesting applications to systems where transport-based methods are less applicable (e.g., insulators). The ability to detect and image both nematicity and spin chirality offers a powerful avenue for investigating multi-\textbf{Q} magnetism, with implications for understanding and engineering topological magnetic states.

\renewcommand{\i}{\ifr}
\let\oldaddcontentsline\addcontentsline% Store \addcontentsline
\renewcommand{\addcontentsline}[3]{}% Make \addcontentsline a no-op

\newpage

\section{Methods}
\label{sec:method}

\subsection{Sample synthesis and characterization}
Single-crystal \CTS{} was synthesized using a standard chemical vapour transport method applied to polycrystalline \CTS{}, as detailed in Refs. \cite{park2022_npj, park2023_ncomm, park2024_composition}. Since the magnetic properties of \CTS{} are sensitive to Co compositions \cite{park2024_composition}, careful assessments of the Co composition were conducted. The \TNt{} value, which reaches a maximum of 26.5\,K with a minimized extent of Co vacancies \cite{park2024_composition}, serves as a reliable indicator of sample quality. Only samples exhibiting \TNt{} = 26.5\,K were used for this study, with an estimated vacancy concentration of less than 3\%.

\subsection{MCD and MLD measurements}

The experimental setup, depicted in Fig.~\ref{fig:intro}b, used wavelength-tunable probe light (typically 650\,nm or 700\,nm) derived from a white light source (xenon lamp) spectrally filtered through a 300\,mm spectrometer. The probe light was mechanically chopped at 137\,Hz, linearly polarized, and then polarization-modulated by a photoelastic modulator (PEM). For MCD, the polarization was modulated between right- and left-circular ($\pm$ quarter-wave modulation) at 50\,kHz. For MLD, the polarization was modulated between linear and cross-linear ($\pm$ half-wave modulation) at 100\,kHz. The \CTS{} samples were mounted in helium vapour in the variable-temperature (2-300\,K) insert of a 7\,T split-coil magnet with direct optical access. The probe light was weakly focused on the sample ($\approx$1~mm spot size) at near-normal incidence, and the reflected light intensity was measured by an avalanche photodiode, amplified, and demodulated using two lock-in amplifiers. We confirmed that perfectly normal incidence ($H \parallel \mathbf{k} \parallel c$) gave the same results. The MCD experiment measured the normalized difference between right- and left-circularly polarized reflected intensities, $(I_R - I_L)/(I_R + I_L)$.  Similarly, MLD measured $(I_\phi - I_{\phi + 90^\circ})/(I_\phi + I_{\phi + 90^\circ})$, where $\phi$ is the angle of the probe's linear polarization. We note that MCD is a close relative of the magneto-optical Kerr effect (MOKE), and in particular the Kerr ellipticity, and is sensitive to magnetic order(s) that generate non-zero off-diagonal conductivity $\sigma_{xy}(\omega)$. Complementing MCD, MLD is sensitive to in-plane anisotropy of the optical conductivity [e.g., $\sigma_{xx}(\omega) - \sigma_{yy}(\omega)$], which can arise from single-\textbf{Q} (stripe-like) AFM order. 

For imaging experiments requiring high spatial resolution, the light source was a 650\,nm superluminescent diode, and the samples were instead mounted on the vacuum cold finger of a small optical cryostat. A high numerical aperture (NA=0.55) microscope objective was used to focus the probe light at normal incidence down to $\approx 1~\mu$m spot, that could be raster-scanned across the sample surface. Small out-of-plane magnetic fields up to 200\,mT were applied using external permanent NdFeB magnets.

\subsection{Classical Monte Carlo simulations}
In addition to the zero-temperature magnetic phase diagram calculation described in Supplementary Note (Section II. E), the phase diagram as a function of temperature and out-of-plane magnetic field ($H$) was obtained using classical Monte Carlo simulations. The simulations employed a combination of the Langevin dynamics algorithm and simulated annealing. All calculations were conducted with the Sunny software package~\cite{sunny_mainref, Sunny}. To achieve statistically well-averaged results, 20 replicas of $36\times 36\times 6$ sized \CTS{} supercell (15,552 Co sites) were prepared and simulated in parallel. The spin systems were initialized by field cooling from 50 K to ensure a uniform alignment of the scalar spin chirality sign, which was necessary to achieve consistent results due to the intertwined nature of the transition between Phases I and II and the chirality sign. At a given temperature and field, we sampled the time evolution of each spin system using the Langevin dynamics after 5,000 Langevin steps for initial thermalization. The Langevin time step and damping constant were set to $\frac{0.05}{S^2(J_{1}+0.1H)}$ (meV$^{-1}$) and 0.1, respectively, where $S=3/2$, $J_{1} = 1.212$\,meV~\cite{park2024_INS}, and $H$ is expressed in units of Tesla. From the collected samples, we calculated the magnitudes of the three Fourier components ($|\tilde{\mathbf{S}}_{\mathbf{Q}_{\nu}}| \equiv \Delta_{\nu}$, see Eq.~\eqref{eq:manifold_main} or Section II. A of Supplementary Note) via Fourier transformation. The results are presented in Fig.~\ref{fig:theory}f--g.

\section{Acknowledgments}
 E.K. and S.A.C. gratefully acknowledge support from the Los Alamos LDRD program and the Quantum Science Center.  P.P. acknowledges support from the U.S. DOE, Office of Science, Basic Energy Sciences, Materials Science and Engineering Division. J.-G.P. acknowledges support from the Samsung Science \& Technology Foundation (Grant No. SSTF-BA2101-05) and the Leading Researcher Program of the National Research Foundation of Korea (Grant No. 2020R1A3B2079375 and RS-2020NR049405). C.D.B. was supported by the U.S. DOE, Office of Science, Office of Basic Energy Sciences, under Award Number DE-SC0022311. The National High Magnetic Field Laboratory is supported by National Science Foundation (NSF) DMR-1644779, the State of Florida, and the U.S. Department of Energy (DOE).  E.K. wants to thank Z. Hawkhead for fruitful discussion.

\let\addcontentsline\oldaddcontentsline% Restore \addcontentsline
\makeatletter
\renewcommand\tableofcontents{%
    \@starttoc{toc}%
}
\makeatother
\renewcommand{\i}{{\rm i}}

%%%%%%%%%%%%%%%%%%%%               SOM               %%%%%%%%%%%%%%%%%%%%%%%%%%%%%%%%%
\onecolumngrid
\vspace{\columnsep}
\begin{center}
\newpage
\makeatletter
{\large\bf{Supplementary Information\\``\@title''}} \smallskip \newline
Erik Kirstein$^1$, Pyeongjae Park$^{2,3}$, Woonghee Cho$^3$, Cristian D. Batista$^{4,5}$, Je-Geun Park$^{3,6}$, Scott A. Crooker,~$^1$\smallskip \\ %\newline
\begin{small}
$^1$National High Magnetic Field Laboratory, Los Alamos National Lab, Los Alamos, New Mexico 87545, USA \\
$^2$Materials Science and Technology Division, Oak Ridge National Laboratory, Oak Ridge, Tennessee 37831, USA \\
$^3$Department of Physics and Astronomy, Seoul National University, Seoul 08826, Korea \\
$^4$Department of Physics and Astronomy, University of Tennessee, Knoxville, TN, USA \\
$^5$Shull Wollan Center—A Joint Institute for Neutron Sciences, \\
Oak Ridge National Laboratory, Oak Ridge, TN, USA \\
$^6$Institute of Applied Physics, Seoul National University, Seoul 08826, Korea \\
\end{small}
\makeatother
\end{center}
\vspace{\columnsep}

% The Supplementary Material includes the following topics:

\twocolumngrid

\renewcommand{\thepage}{S\arabic{page}}
\renewcommand{\theequation}{S\arabic{equation}}
\renewcommand{\thefigure}{S\arabic{figure}}
\renewcommand{\bibnumfmt}[1]{[#1]}
\renewcommand{\citenumfont}[1]{#1}

\setcounter{page}{1}
\setcounter{section}{0}
\setcounter{equation}{0}
\setcounter{figure}{0}

\section{Experimental details}

\subsection{Spectral dependence of MCD and MLD optical response}

The magnitude and sign of the MCD signals that arise from the  spin chirality of the noncoplanar triple-$\mathbf{Q}$ antiferromagnetism in Co$_{1/3}$TaS$_2$ depend on the wavelength $\lambda$ -- or equivalently, the photon energy -- of the probe light. Figure S1a shows examples of MCD versus applied out-of-plane magnetic field $H$ hysteresis loops, taken with probe light having different $\lambda$, at a fixed temperature $T$=15~K. The amplitude of the open hysteresis loop varies with $\lambda$, peaking at $\lambda \approx 700$\,nm and inverting sign around $\lambda \approx 950$\,nm.  Moreover, the slope of the linear background is also $\lambda$-dependent.  And finally, the magnitudes of the additional steps that appear at the metamagnetic transition at $H_m \approx \pm3.5$~T also vary with $\lambda$ (this latter aspect is discussed further in the next section, which considers offsets and artifacts in MCD measurements that can arise from linear dichroism). 

\begin{figure*}[tbp]
\centering
\includegraphics[width=1.8\columnwidth]{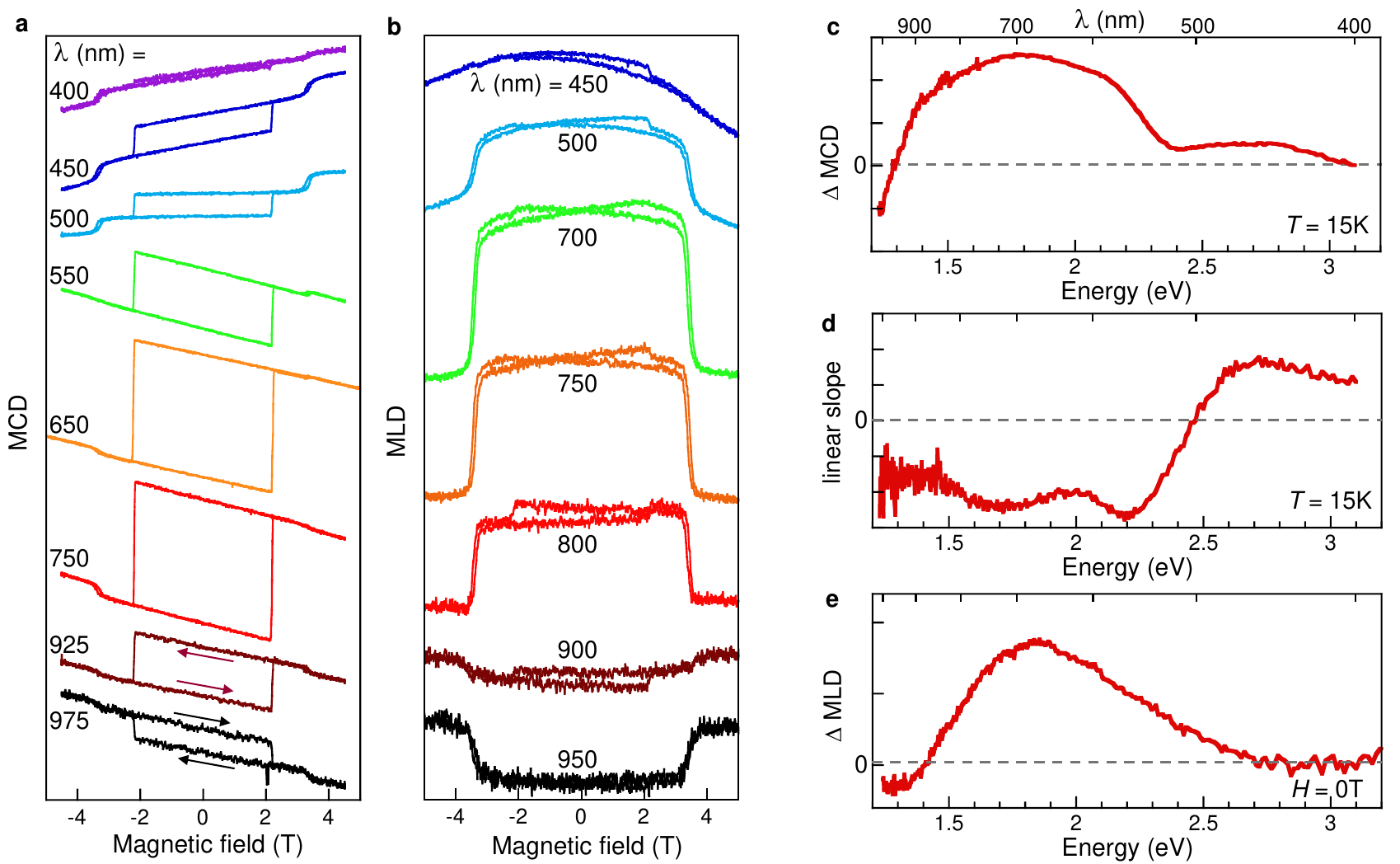}
\caption{\textbf{Dependence of MCD and MLD signals on the wavelength $\lambda$ of the probe light.} \textbf{a}, Examples of MCD($H$) hysteresis loops taken on Co$_{1/3}$TaS$_2$ at a fixed temperature (15~K), using probe light of different wavelengths (curves vertically offset for clarity). The amplitude (and sign) of the open hysteresis loop, the slope of the linear background, and the amplitude of the steps at the metamagnetic transition $H_m \approx \pm 3.5$~T, all vary with $\lambda$. \textbf{b}, Examples of MLD($H$) measurements taken on Co$_{1/3}$TaS$_2$ at a fixed temperature (15~K), using probe light of different wavelengths. The magnitude and sign of the MLD signals that emerge for $|H|<H_m$ vary with $\lambda$. Note that a parabolic background is present at short $\lambda$. \textbf{c-e}, Explicit spectral dependencies, plotted as a continuous function of the probe light's photon energy, of \textbf{c}, the amplitude of the open hysteresis loop in MCD, \textbf{d}, the background linear slope of the MCD, and \textbf{e}, the amplitude of the MLD signal. These three spectral dependencies, which are not the same, derive from the interband optical transitions within the band structure, which are influenced by the underlying AFM spin order in Co$_{1/3}$TaS$_2$. These detailed spectral dependencies can be used to benchmark first-principles calculations of the diagonal and off-diagonal complex optical conductivities $\sigma_{xx}(\omega)$, $\sigma_{yy}(\omega)$, and $\sigma_{xy}(\omega)$ at optical frequencies.} 
\end{figure*}

Similarly, the MLD signals that arise from the single-$\mathbf{Q}$ (stripe-like) AFM order in Co$_{1/3}$TaS$_2$ are also $\lambda$-dependent, as can be seen from the examples of MLD($H$) curves shown in Fig. S1b. 

These various spectral dependencies were explicitly measured as a continuous function of probe wavelength $\lambda$, and are plotted in Figs.~S1c-e as a function of the photon energy of the probe light.  The amplitude of the open hysteresis loop in MCD studies was determined via the difference of two MCD($\lambda$) spectra acquired at $H=+1.5$~T, one acquired after ramping $H$ down from large positive value, and the other acquired after ramping $H$ up from large negative value (thus, on the upper and lower branches of the hysteresis loop). The amplitude of the MLD signal was derived via the difference of two zero-field MLD($\lambda$) spectra, one acquired at $T=40$~K (in the nonmagnetic state) and the other acquired at $T=27$~K (in the single-$\mathbf{Q}$ Phase III). 

Each of these spectral dependencies is different and has its own ``spectral fingerprint'', as expected because each of these quantities derives from different combinations of the diagonal and off-diagonal complex conductivities $\sigma_{xx}(\omega)$, $\sigma_{yy}(\omega)$, and $\sigma_{xy}(\omega)$ at optical frequencies. In turn, these optical conductivities derive from the underlying spin order in Co$_{1/3}$TaS$_2$ and the influence of the spin order on the band structure and on the interband optical transitions occurring within this band structure. Given that spin-polarized bandstructures and frequency-dependent optical conductivities $\sigma(\omega)$ are routinely calculated for magnetic materials by a variety of first-principles theoretical methods (e.g., density functional theory), we anticipate that broadband  magneto-optical measurements of MCD and MLD (and also closely-related effects such as Kerr rotation and Kerr ellipticity) can provide important benchmarks against which  theoretical approaches can be compared and refined.

\subsection{Cross-talk between measurements of circular and linear dichroism}
 
\begin{figure*}[tbp]
\centering
\includegraphics[width=1.7\columnwidth]{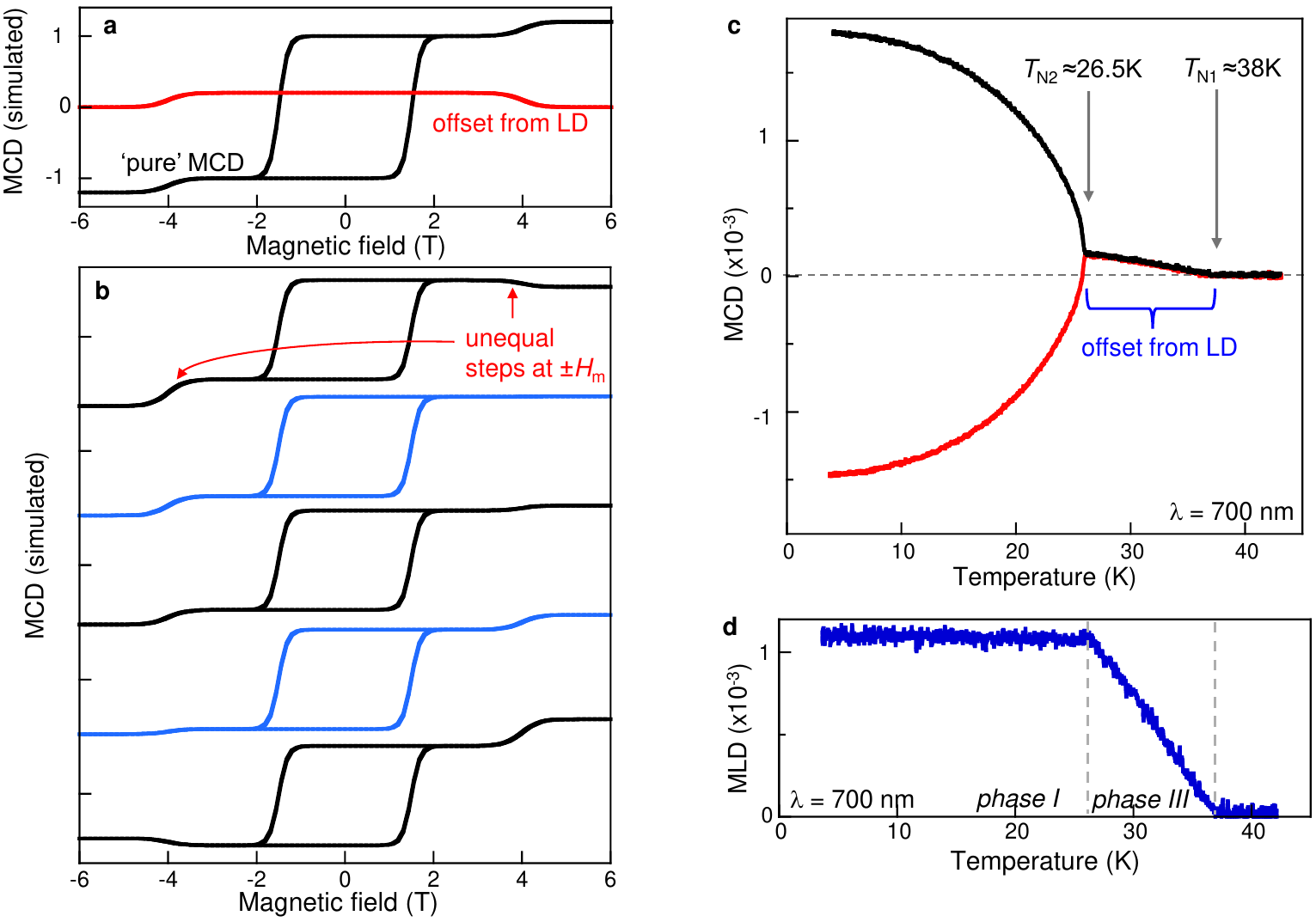}
\caption{\textbf{Cross-talk between MCD and MLD signals.} The presence of linear dichroism in a sample (due, e.g., to the emergence of nematic order) can lead to offsets (artifacts) in measurements of circular dichroism.  \textbf{a}, Black trace: a simulated ``pure'' MCD($H$) signal from chiral triple-$\mathbf{Q}$ AFM order in Co$_{1/3}$TaS$_2$, showing equal changes at the metamagnetic transitions at $\pm H_m$ ($= \pm$4\,T in these simulations).  Red trace: A small offset due to linear dichroism (simulated), that appears when $|H|<H_m$. \textbf{b}, Adding or subtracting various amounts of  this offset to the pure MCD($H$) curve results in MCD($H$) traces that have \textit{unequal} step sizes at $H= \pm H_m$, as experimentally observed (\textit{cf.} Figure S1).  \textbf{c}, Cross-talk between MCD and MLD signals can also be seen in some MCD($T$) data, as a small linear change in the MCD signal in the temperature range between 38~K and 26.5~K (\textit{i.e.}, in Phase III), where only single-\textbf{Q} nematic order (and therefore only linear dichroism) emerges in Co$_{1/3}$TaS$_2$. The black and red curves were acquired while warming from $T = 4 \rightarrow 43$~K in $H$=0, after cooling to $T$=4~K in $H=\pm 0.5$~T respectively (which poles the  chirality of the non-equilateral triple-\textbf{Q} AFM order in Phase I). \textbf{d}, For reference, the measured MLD($T$) at the same location is shown, showing the emergence of linear dichroism due to single-\textbf{Q} AFM order upon cooling from 38~K to 26.5~K}  
\end{figure*}

These experiments use a commercial photoelastic modulator (PEM) from Hinds Instruments to modulate the polarization of the probe light between right- and left-circular for MCD studies (and between linear and cross-linear for MLD studies). In measurements of this type, it has long been known~\cite{Shindo1990, Ugras2023} that the presence of any \textit{linear} dichroism (or linear birefringence) in the sample, can generate small artifacts and offsets during measurements of \textit{circular} dichroism (and vice-versa).  In other words, changes in a material's linear dichroism (due, e.g., to the emergence of nematic order and concomitant anisotropy of the in-plane optical conductivity) can lead to offsets and artifacts in measurements of circular dichroism.  In our measurements of Co$_{1/3}$TaS$_2$, signatures of these artifacts or ``cross-talk'' between circular dichroism and linear dichroism measurements are apparent in the unequal steps of the MCD($H$) signals at the metamagnetic phase transition fields $H_m \approx \pm 3.5$~T. For $|H|>H_m$ there is no linear dichroism (i.e., no nematic order), but for $|H|<H_m$ linear dichroism exists (because single-\textbf{Q} magnetism and nematicity are present), which causes a small offset in the MCD signal in this low-field range.  

As shown pictorially in Fig. S2, this effect causes the steps in MCD at $+H_m$ and $-H_m$ to be unequal and no longer symmetric.  As discussed in the previous section, the amount of cross-talk is wavelength-dependent, as can be seen from the various MCD($H$) curves in Fig. S1a. Note that at wavelengths where the MLD signals due to nematicity are approximately zero (i.e., at $\lambda \approx$ 450~nm and 900~nm), the offsets vanish and MCD($H$) traces have the expected symmetry -- meaning that the steps in MCD at $\pm H_m$ are approximately equivalent.

Further, we emphasize that both the magnitude and sign of the cross-talk can also depend on sample location:  If the probe beam interrogates a region of the sample containing nematic domains oriented primarily along (for example) 0$^\circ$, then the offset due to cross-talk will have a particular sign (e.g., positive).  But if the probe beam is moved to a different region of the sample where the nematic domains are oriented primarily along 120$^\circ$ or 240$^\circ$, then the offset due to cross-talk will have opposite (negative) sign.

A related manifestation of this cross-talk can be seen in some measurements of MCD vs temperature, where the MCD signal can appear to increase in magnitude slowly and linearly as the sample cools from 38~K ($T_{N1}$) down to 26.5~K ($T_{N2}$). A discussed above, this is simply a consequence of the emerging linear dichroism due to single-$\mathbf{Q}$ AFM order, causing a small offset in the measurement of circular dichroism. An example of this behavior can be seen in the MCD($T$) data shown in Fig. S2c. 

\subsection{Mapping out the magnetic phase diagram of Co$_{1/3}$TaS$_2$: Temperature-dependent MLD($H$)}

\begin{figure}[tbp]
\centering
\includegraphics[width=0.95\columnwidth]{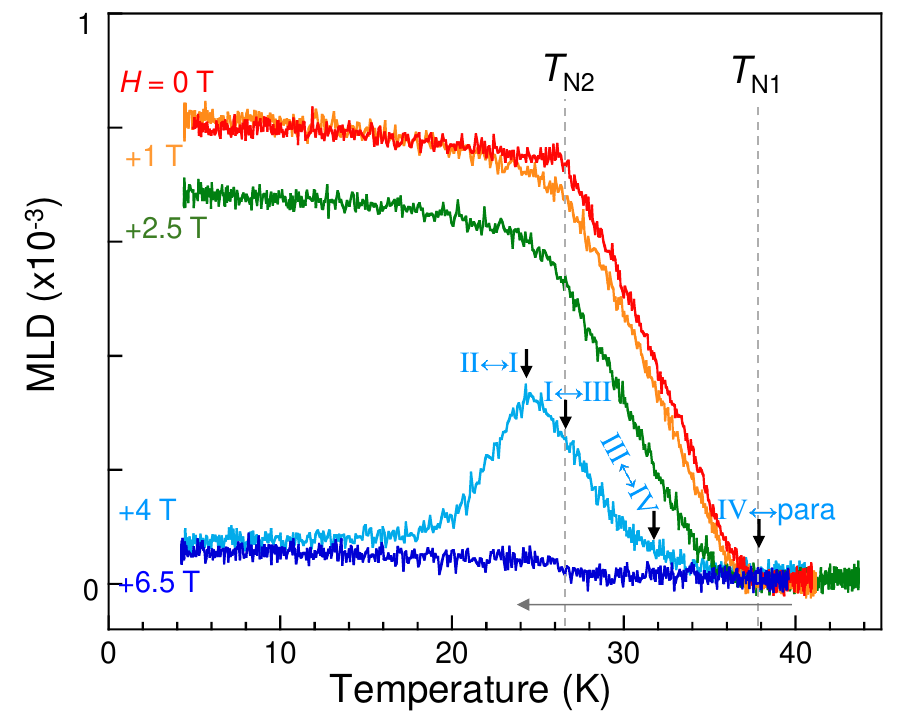}
\caption{\textbf{Temperature-dependent MLD at different applied fields $H$.}  MLD($T$) data help to map out the boundaries of the Co$_{1/3}$TaS$_2$ magnetic phase diagram, and in particular to identify transitions between phases with different nematic order. At $H$=0~T (red curve), the transition from the paramagnetic state to the single-$\mathbf{Q}$ nematic (stripe) phase III is evident at $T_{N1}$, as is the abrupt saturation of the MLD signal upon further cooling into phase I at $T_{N2}$.  In contrast, at $H$=+6.5~T (dark blue), the MLD remains approximately zero throughout cooling, indicating that both high-field phases II and IV lack nematic order. Notably, cooling at $H$=+4T (light blue curve) crosses all four magnetic phases IV$\rightarrow$III$\rightarrow$I$\rightarrow$II, which highlights the utility of the continuous multi-$\mathbf{Q}$ framework.  All data taken using 650~nm probe light. }  
\end{figure}

To map out the magnetic phase diagram of Co$_{1/3}$TaS$_2$, we measured not only $H$-dependent MCD and MLD curves at different fixed temperatures (shown in Fig. 2 of the main text), but also measured temperature-dependent MCD and MLD curves at different fixed $H$. Effectively, these represent horizontal line cuts through the ($H,T$) phase diagram shown in Fig. 2 of the main text.   

Examples of MLD($T$) curves are shown in Fig. S3, acquired while slowly cooling the sample from approximately $40 \rightarrow 4$~K, at $H$=0, +1, +2.5, +4, and +6.5~T.  As described in the figure caption, these data help to identify phase boundaries at which the single-$\mathbf{Q}$ nematic (stripe) order changes.  Especially noteworthy is the MLD($T$) curve acquired at $H$=+4~T, where the sample goes through all four magnetic phases of Co$_{1/3}$TaS$_2$ as it cools (paramagnetic $\rightarrow$ IV $\rightarrow$ III $\rightarrow$ I $\rightarrow$ II).

\subsection{Additional MLD images and analysis of single-Q domains}

\begin{figure}[h]
\centering
\includegraphics[width=0.9\columnwidth]{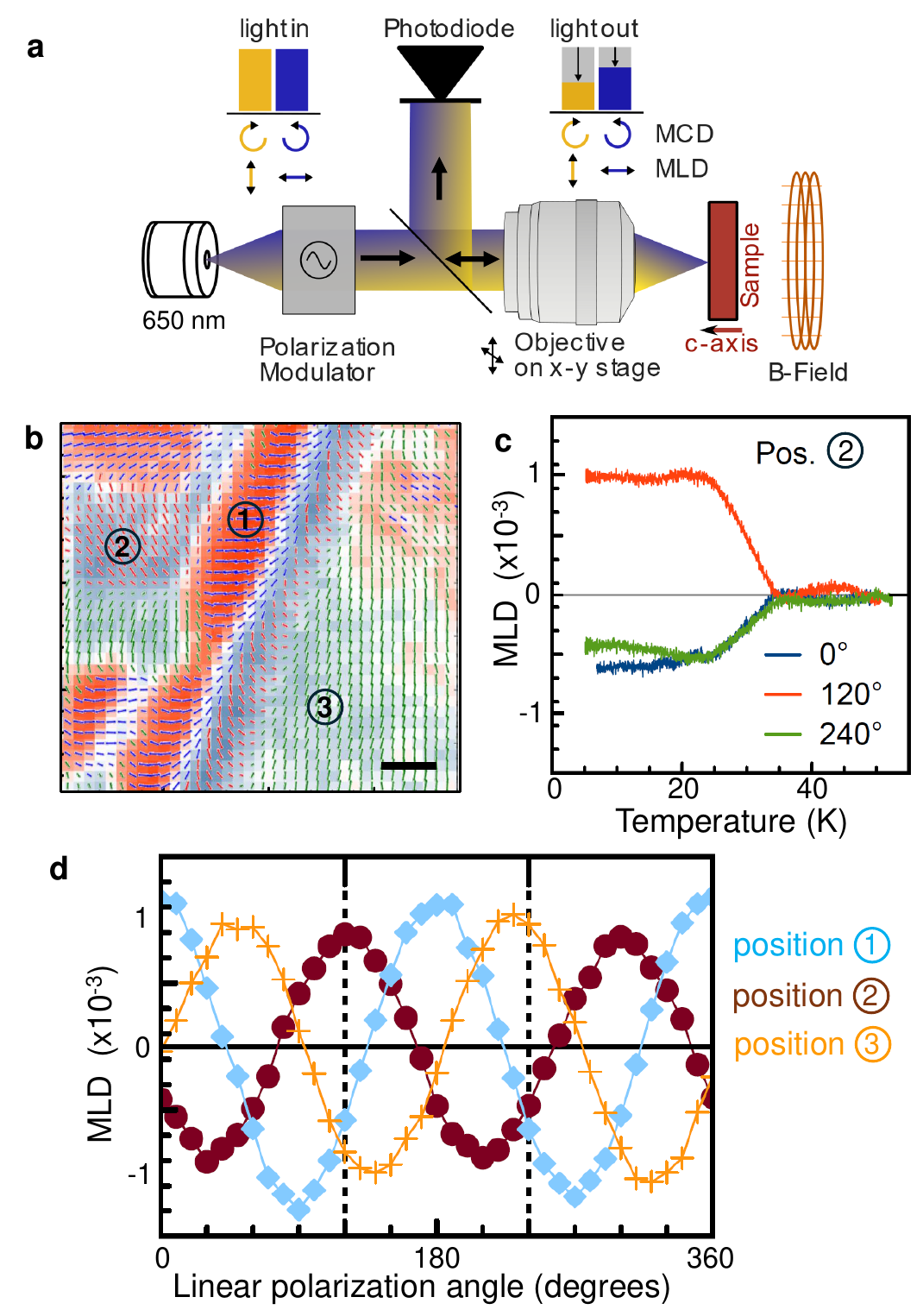}
\caption{\textbf{Characterizing single-Q AFM domains with MLD microscopy}. \textbf{a}, Schematic of the microscopy setup. \textbf{b}, Image of nematic director field superimposed on an MLD image taken using probe light linearly polarized along $\phi = 0^\circ$ (same image as Fig 4d of main text). \textbf{c}, Temperature-dependent MLD taken at position 2 (where the nematic director is oriented primarily along $120^\circ$), using probe light linearly polarized along $\phi = 0^\circ, 120^\circ, 240^\circ$. \textbf{d}, MLD signals taken at the three positions indicated, versus the probe light's linear polarization angle $\phi$. To remove background signals that are independent of magnetism, these data are the \textit{difference} of MLD($\phi$) scans taken in the single-$\mathbf{Q}$ state ($\approx$\,27\,K) and in the nonmagnetic state ($\approx$\,38\,K). These data were used to construct the polar plots shown in Fig. 4b of the main text.}  
\end{figure}

Figure S4 shows a further analysis of the MLD signals and images that were presented in Fig. 4 of the main text. Fig. S4b shows the nematic director field superimposed on an MLD image (same as Fig. 4d of the main text), with three specific locations indicated. Fig. S4c shows temperature-dependent MLD scans at location \#2 (where the nematic director field is oriented along an angle $\theta_{1Q} \approx 120^\circ$), using probe light that is linearly polarized along $\phi = 0^\circ, 120^\circ$, and $240^\circ$. When $\phi = 120^\circ$, the MLD signal increases by a large amount in the positive direction upon cooling from the nonmagnetic state ($\sim$40~K) into the single-$\mathbf{Q}$ AFM phase III (similar, e.g., to Fig. 1g in the main text). This is because $\phi$ is aligned along the nematic director $\theta_{1Q}$ at this location.  In contrast, repeating this temperature-dependent MLD scan using light polarized along $\phi = 0^\circ$ or $240^\circ$, the MLD increases in the \textit{negative} direction, but only by about half the amount. This is in line with, and validates, the expectation that MLD signals should scale as $1/2 - \textrm{sin}^2(\phi - \theta_{1Q})$.

Figure S4d shows measurements of MLD versus $\phi$ at each of the three locations indicated. Each shows a $\textrm{sin}(2\phi)$ dependence, as expected for MLD, but shifted in phase by $\pm 120^\circ$. These data were used to construct the polar plots shown in Fig. 4 of the main text. 

\begin{figure*}[tbp]
\centering
\includegraphics[width=1.8\columnwidth]{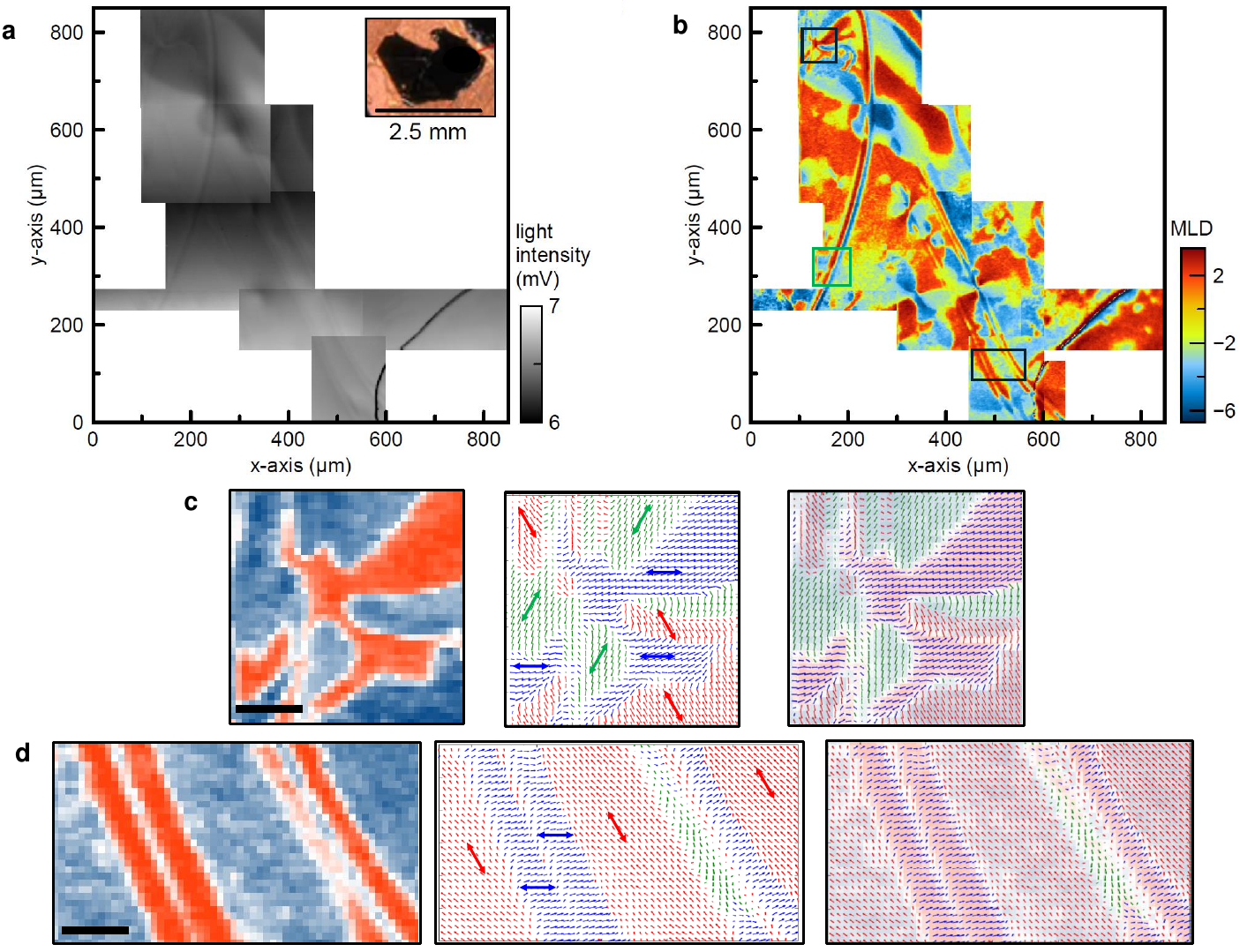}
\caption{\textbf{Spatially-resolved MLD images over a larger area}. \textbf{a}, A composite image showing the intensity of the reflected probe light used in MLD imaging studies, over different regions on the Co$_{1/3}$TaS$_2$ sample surface, from which surface imperfections can be seen (note: a slow linear drift of the reflected intensity within each sub-image was subtracted). \textbf{b}, The corresponding image of the MLD signal. Comparison with panel \textbf{a} suggests that single-$\mathbf{Q}$ AFM domains may be pinned by structural features and faults, and associated local strain fields.  \textbf{c}, $70 \times 70~\mu$m MLD image of a different region of the Co$_{1/3}$TaS$_2$ sample, indicated by the black square in panel \textbf{b}, acquired using linearly-polarized light modulated between 0$^\circ$ and 90$^\circ$. The measured nematic director map is shown to the right, and on the far right is the director map superimposed on the MLD image. \textbf{d}, Same, but for a different $110 \times 60~\mu$m region, indicated by the black rectangle in panel \textbf{b}. Note: The green square in panel \textbf{b} indicates the region imaged in Fig. 4 of the main text.  All data taken using 650~nm probe light. Scale bars are 20\,$\mu$m. }  
\end{figure*}

Additional MLD images were also acquired over a larger area of the sample surface, spanning nearly 1~mm $\times$ 1~mm. Figure S5a shows a composite image of the reflected intensity of the probe light used to measure the MLD, over several rectangular regions of the sample surface. From these images, slight variations in the reflected intensity can be observed, revealing what are likely structural imperfections and faults. The corresponding MLD images are shown in Fig. S5b, where it is readily seen that MLD domains are often co-located with the structural imperfections, suggesting that local strain fields likely play a role in determining the patterns of single-$\mathbf{Q}$ AFM domains. These MLD images were acquired at 8~K,  and were not corrected for any background linear dichroism in the paramagnetic state (that is, images acquired at $T>T_{N1}$ were not subtracted).

Figures S5c and S5d show high-resolution images of the measured MLD and associated nematic director fields at two other regions of the Co$_{1/3}$TaS$_2$ sample surface (indicated by black rectangles  in Fig. S5b). As discussed in the main text, the nematic domains are large (many tens of microns) and are found to align primarily along the 0$^\circ$, 120$^\circ$, and 240$^\circ$ directions.

\section{Theoretical model for the continuous multi-\texorpdfstring{$\mathbf{Q}$}{Q} manifold of \texorpdfstring{$\mathbf{M}$}{M}-ordering}
In this section, we present a comprehensive theoretical model that describes the general $M$-orderings on a triangular lattice, i.e., antiferromagnetic orderings where the ordering wave vectors $\mathbf{Q}_{\nu} (\nu=1,2,3)$ equal half the reciprocal lattice wave vectors and therefore correspond to the $M$ points of the hexagonal Brillouin zone. To capture the rich phase diagram suggested in Fig.~2c, it is essential to consider both four-spin interactions and magnetic anisotropy, in addition to bilinear Heisenberg interaction terms. We elaborate on the role of each term in activating the continuous multi-$\mathbf{Q}$ manifold and show how the resultant spin model accounts for key experimental observations in \CTS{}.

\begin{figure*}[ht]
  \includegraphics[width=1\linewidth]{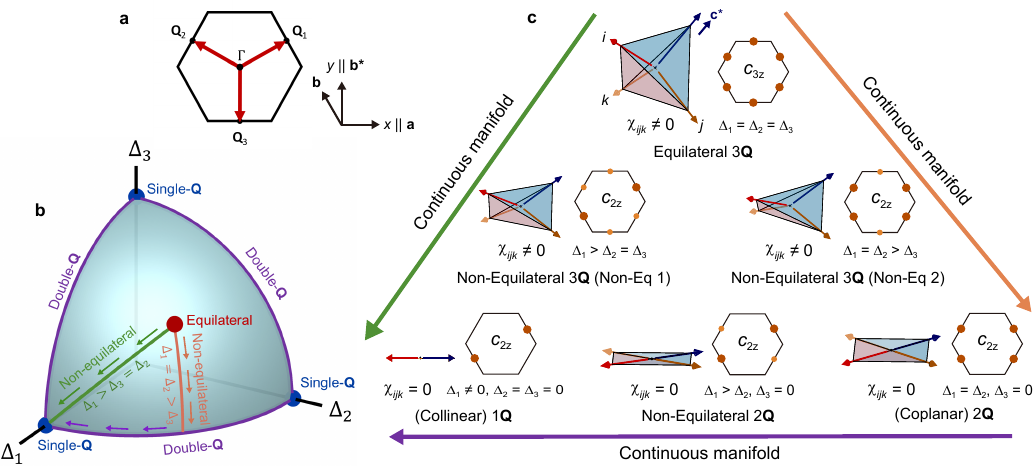}%\\
  \caption{Continuous manifold of multi-$\mathbf{Q}$ $M$-ordering spanned by Eq.~\eqref{eq:manifold}. \textbf{a}, Real and reciprocal lattice coordinate conventions referenced in this study. The three red arrows denote the ordering wave vectors of the $M$-ordering, $\mathbf{Q}_{\nu}$ ($\nu$=1,2,3). \textbf{b}, Continuous manifold of multi-$\mathbf{Q}$ $M$-orderings, shown as a variational space on a spherical shell. Same as Fig.~\ref{fig:theory}a but with an additional highlight of the orange path connecting the equilateral triple-$\mathbf{Q}$ and double-$\mathbf{Q}$ orderings. \textbf{c}, Magnitude of the Fourier components ($\Delta_{\nu}$) and the corresponding real-space configurations of the four spin sublattices for various phases on the manifold.}
\label{Sfig:multiQ}
\end{figure*}

Our mathematical formulation is based on the real and reciprocal space coordinates shown in Fig.~\ref{Sfig:multiQ}a, which leads to $\mathbf{Q}_{1} = \mathbf{a}^{*}/2$, $\mathbf{Q}_{2} = -\mathbf{a}^{*}/2 + \mathbf{b}^{*}/2$, and $\mathbf{Q}_{3} = -\mathbf{b}^{*}/2$. In other words, $\mathbf{Q}_{\nu} = \mathbf{G}_{\nu}/2$ with $\nu=1,2,3$, where $\mathbf{G}_{\nu}$ are reciprocal lattice vectors related by three-fold rotation about the $c$-axis.

To describe antiferromagnetic orderings characterized by specific ordering wave vectors, it is convenient to express the spin Hamiltonian in momentum space. We begin with the most general isotropic Hamiltonian that includes bilinear Heisenberg and four-spin interactions:
\begin{eqnarray}
{\hat{\cal H}} = {\hat{\cal H}}_{\rm Heis} + {\hat{\cal H}}_{\rm K}
\end{eqnarray}
with
\begin{eqnarray}
{\hat{\cal H}_{\rm Heis}} &=& \sum_{ \mathbf{q}} \tilde{J}^{ab}_{\mathbf{q}} \tilde{\mathbf{S}}^a_{\mathbf{q}} \cdot  \tilde{\mathbf{S}}^b_{-\mathbf{q}} \nonumber \\
\hat{\cal H}_{\rm K} &=& \!\!\!\!\! \sum_{\mathbf{q}_{i},\mathbf{q}_{j},\mathbf{q}_{k}} 
\!\!\!\!\! \frac{\tilde{K}^{abcd}_{\mathbf{q}_i\mathbf{q}_j\mathbf{q}_k}}{N} (\tilde{\mathbf{S}}^a_{\mathbf{q}_{i}} \cdot  \tilde{\mathbf{S}}^b_{\mathbf{q}_{j}})(\tilde{\mathbf{S}}^c_{\mathbf{q}_{k}} \cdot  \tilde{\mathbf{S}}^d_{-\mathbf{q}_{\mathrm{tot}}}),
\label{eq:Hgeneral}
\end{eqnarray}
where we are adopting the convention of summation for repeated sublattice indices $a,b,c,d \in \{o,e \}$ with $o, e$ denoting odd and even layers of two Co sublattices, $\mathbf{q_{\mathrm{tot}}} = \mathbf{q}_{i}+\mathbf{q}_{j}+\mathbf{q}_{k}$ and  
\begin{equation}
\tilde{\mathbf{S}}^a_{\mathbf{q}}  = \frac{1}{\sqrt{N}} \sum_{\mathbf{r}} e^{-i\mathbf{q}\cdot\mathbf{r}}\mathbf{S}^a_{\mathbf{r}} 
\end{equation}
is the Fourier transform of the spin field, where $\mathbf{r}$ spans $N$ crystallographic  unit cells of the material of interest. 

In this analysis, we assume that ${\hat{\cal H}}_{\rm Heis}$ dominates over ${\hat{\cal H}}_{\rm K}$, a hierarchy expected to hold in most real materials, including \CTS{} \cite{park2023_ncomm, park2024_INS}. Under this assumption, we demonstrate how the general spin model in Eq.~\eqref{eq:Hgeneral} simplifies for the   $M$-ordering problem in \CTS{}.

\subsection{Bilinear Heisenberg interactions and degenerate multi-\texorpdfstring{$\mathbf{Q}$}{Q} manifold}
The first step in analyzing Eq.~\eqref{eq:Hgeneral} is to realize $M$-ordering by restricting the bilinear Heisenberg interactions so that they favor magnetic ordering wave vectors $\mathbf{Q}_{\nu} = \mathbf{G}_{\nu}/2$ [Fig.~\ref{Sfig:multiQ}a]. For isotropic Heisenberg interactions described by ${\hat{\cal H}}_{\rm Heis}$ this implies that $\tilde{J}_{\mathbf{q}}$ has global minima at $\mathbf{q} = {\mathbf{Q}}_{\nu}$ with $\nu= 1, 2, 3$ (see Fig.~\ref{Sfig:multiQ}a). When this condition holds, the classical ground state of the system has nonzero Fourier components $\tilde{\mathbf{S}}_{\mathbf{q}}$ only at  $\mathbf{q}=\mathbf{Q}_{\nu}$. Additionally, a finite Fourier component at $\mathbf{q}=\mathbf{0}$--representing non-zero magnetization--is expected to be induced by a uniform magnetic field.

The most general real-space spin configuration that minimizes ${\hat{\cal H}}_{\rm Heis}$ described above can be expressed as
\begin{equation}
{\mathbf{S}}^a({\mathbf{r}}) = \sum_{\nu} \tilde{\mathbf{S}}^a_{{\mathbf{Q}}_{\nu}} \cos({{\mathbf{Q}}_{\nu} \cdot {\mathbf{r}}}),
\label{eq:manifold}
\end{equation}
where ${\mathbf{r}}$ is a lattice vector on the two-dimensional triangular lattice; the out-of-plane component of the ordering wave vectors $\mathbf{Q}_{\nu}$ is equal to zero [Fig.~\ref{Sfig:multiQ}a]. 

For the two Co sublattices arranged in a hexagonal close-packed structure [see Fig.~\ref{fig:intro}a], the antiferromagnetic exchange interactions between even and odd layers \cite{park2023_ncomm, park2024_INS} ensure that the spin configuration in one layer matches the other when shifted by the vector $\mathbf{t} = (1, 1, 1/2)$ in lattice units. Consequently, the vector amplitudes for the even and odd sublattices are connected by the following simple relation:
\begin{equation}
\tilde{\mathbf{S}}^{o}_{\mathbf{Q}_{\nu}} = e^{i {\bf Q}_{\nu} \cdot {\mathbf{t}}}\, \tilde{\mathbf{S}}^{e}_{\mathbf{Q}_{\nu}}.
\label{eq:sublattice}
\end{equation}
Given this relationship, in the following we will drop the sublattice index: $\tilde{\mathbf{S}}^{o}_{\mathbf{Q}_{\nu}} \equiv \tilde{\mathbf{S}}_{\mathbf{Q}_{\nu}}$
and $\tilde{\mathbf{S}}^{e}_{\mathbf{Q}_{\nu}} \equiv e^{i {\bf Q}_{\nu} \cdot {\mathbf{t}}} \tilde{\mathbf{S}}_{\mathbf{Q}_{\nu}} $. In other words, the identical spin configurations across adjacent Co layers effectively reduce the 3D magnetic model of \CTS{} to that of a 2D triangular magnetism. We define $|\tilde{\mathbf{S}}^o_{\mathbf{Q}_{\nu}}|= |\tilde{\mathbf{S}}^e_{\mathbf{Q}_{\nu}}| \equiv \Delta_{\nu}$ and will use $\Delta_{\nu}$ from hereafter.

Given that the system exhibits the shortest antiferromagnetic periodicity ($\mathbf{Q}_{\nu} = \mathbf{G}_{\nu}/2$), Eq.~\eqref{eq:manifold} allows for a maximum of four spin sublattices in real space. Consequently, the problem further reduces to a four-site system with real-space coordinates:
\begin{eqnarray}
{\mathbf{r}}_1 = {\mathbf{0}}, \quad {\mathbf{r}}_2 = {\mathbf{a}}, \quad {\mathbf{r}}_3 = {\mathbf{b}}, \quad  {\mathbf{r}}_4 = {\mathbf{a}} + {\mathbf{b}},
\end{eqnarray}
where ${\mathbf{a}}$ and ${\mathbf{b}}$ are primitive lattice vectors (see Fig.~\ref{Sfig:multiQ}a). The four nonzero Fourier components at $\mathbf{q}=\mathbf{0}$ and $\mathbf{q}=\mathbf{Q}_{\nu}$ -- with the first case corresponding to a ferromagnetic component -- are related to the four classical spin vectors at $\mathbf{r}_{j}~(j=1,2,3,4)$ as follows:

\begin{eqnarray}
\tilde{\mathbf{S}}_{\mathbf{0}} &=& \frac{\sqrt{N}}{4} ({\mathbf{S}}_{1} + {\mathbf{S}}_{2} + {\mathbf{S}}_{3}+ {\mathbf{S}}_{4})
\nonumber \\
\tilde{\mathbf{S}}_{{\mathbf{Q}}_1} &=& \frac{\sqrt{N}}{4} ({\mathbf{S}}_{1} - {\mathbf{S}}_{2} + {\mathbf{S}}_{3} -{\mathbf{S}}_{4})
\nonumber \\
\tilde{\mathbf{S}}_{{\mathbf{Q}}_2} &=& \frac{\sqrt{N}}{4} ({\mathbf{S}}_{1} - {\mathbf{S}}_{2} - {\mathbf{S}}_{3} +{\mathbf{S}}_{4})
\nonumber \\
\tilde{\mathbf{S}}_{{\mathbf{Q}}_3} &=& \frac{\sqrt{N}}{4} ({\mathbf{S}}_{1} + {\mathbf{S}}_{2} - {\mathbf{S}}_{3} -{\mathbf{S}}_{4}),
\label{eq:transform}
\end{eqnarray}
where $\tilde{\mathbf{S}}_{\mathbf{0}}$ is proportional to the net spin magnetization.

In the classical limit, the four spins $\mathbf{S}_{{i}}$ have a fixed magnitude, $|\mathbf{S}_{{i}}| = S$. While the longitudinal spin stiffness remains finite in magnetic systems with sizable quantum spin fluctuations (which usually manifest in small-spin systems), this classical approximation should adequately describe the ground-state ordering and its long-wavelength spin fluctuations in \CTS{}, where $S = 3/2$ arises from the $d^7$ high-spin configuration of $\text{Co}^{2+}$. For compensated antiferromagnetic configurations, where $\tilde{\mathbf{S}}_{\mathbf{0}} = \mathbf{0}$, this spin-length constraint is satisfied when the three vector Fourier components $\mathbf{S}_{\mathbf{Q}_{\nu}}$ are mutually orthogonal,
\begin{equation}
\tilde{\mathbf{S}}_{\mathbf{Q}_{\nu}} \perp \tilde{\mathbf{S}}_{\mathbf{Q}_{\nu'}} \,\,\,\mathrm{for}\,\,\, \nu \neq \nu'
\label{eq:ortho}
\end{equation}
and satisfy the normalization constraint:
\begin{equation}
\sum_{\nu=1}^{3} \tilde{\mathbf{S}}_{{\mathbf{Q}}_{\nu}} \cdot  \tilde{\mathbf{S}}_{{\mathbf{Q}}_{\nu}} = \sum_{\nu} {{\Delta_{\nu}}^2} = NS^2. 
\label{eq:sumrule}
\end{equation}

Under the constraints of Eq.~\eqref{eq:ortho} and \eqref{eq:sumrule}, all multi-$\mathbf{Q}$ spin configurations generated by Eq.~\eqref{eq:manifold} are confined to a spherical surface spanned by the three orthogonal basis vectors $\tilde{\mathbf{S}}_{\mathbf{Q}_{\nu}}$. As illustrated in Fig.~\ref{fig:theory}a or Fig.~\ref{Sfig:multiQ}b, this schematic in the $\Delta_{1}-\Delta_{2}-\Delta_{3}$ phase space provides a clear representation of the multi-$\mathbf{Q}$ manifold explored in this work. By varying the relative magnitudes of $\Delta_{\nu}$, one can access all possible single-, double-, and triple-$\mathbf{Q}$ spin configurations corresponding to the $M$-ordering.

Figs.~\ref{Sfig:multiQ}b--c show two representative high-symmetry paths on this multi-$\mathbf{Q}$ manifold, describing the continuous interpolation between the most symmetric triple-$\mathbf{Q}$ ordering and the single-$\mathbf{Q}$ ordering (green or orange/purple arrows). The state with $\Delta_1 = \Delta_2 = \Delta_3$ corresponds to the equilateral tetrahedral triple-$\mathbf{Q}$ ordering, where the four spin sublattices individually align with the principal directions of a regular tetrahedron. Notably, this is the only configuration that preserves the hexagonal rotation symmetry of the triangular lattice. However, when $\Delta_1 = \Delta_2$ but $\Delta_3 \neq \Delta_1$, the tetrahedron spanning the four spin sublattices becomes non-equilateral (labeled Non-Eq 1 or Non-Eq 2 in Fig.~\ref{Sfig:multiQ}c). These intermediate states smoothly connect the equilateral triple-$\mathbf{Q}$ ordering to either a single-$\mathbf{Q}$ or double-$\mathbf{Q}$ ordering. Importantly, these intermediate states exhibit two-fold rotational symmetry ($C_{2z}$), similar to single-$\mathbf{Q}$ or double-$\mathbf{Q}$ orderings, while retaining the net scalar spin chirality of the equilateral triple-$\mathbf{Q}$ ordering, consistent with the symmetry argument for Phase I in Fig.~\ref{fig:hysttemperature}c.

It is important to emphasize that, under the constraints of Eq.~\eqref{eq:ortho} and \eqref{eq:sumrule}, and given that $\mathbf{Q}_\nu = -\mathbf{Q}_\nu$, the continuous multi-$\mathbf{Q}$ manifold spanned by Eq.~\eqref{eq:manifold} remains fully degenerate when ${\hat{\cal H}} = {\hat{\cal H}}_{\rm Heis}$. Additionally, we stress that magnetic anisotropy alone cannot induce a multi-$\mathbf{Q}$ ground state, as will be discussed in the following sections. Therefore, the four-spin term ${\hat{\cal H}}_{\rm K}$ is essential for realizing four-sublattice multi-$\mathbf{Q}$ magnetic ground states, as also highlighted in Refs.~\cite{batista2016_review, park2023_ncomm, park2024_INS}.

\subsection{Four-spin interactions}

The accidental multi-$\mathbf{Q}$ degeneracy described earlier is lifted by $\hat{\cal H}_{\rm K}$. As noted in the introduction to Section II, $\hat{\cal H}_{\rm K}$ is much weaker than ${\hat{\cal H}}_{\rm Heis}$, ensuring that it does not generate any additional Fourier components beyond the four $\tilde{\mathbf{S}}_{{\mathbf{Q}}_{\nu}} (\nu=0,1,2,3)$ in Eq.~\eqref{eq:transform}. 
Thus, by using Eq.~\eqref{eq:sublattice}, the classical energy due to four-spin interactions [$\hat{\cal H}_{\rm K}$ in Eq.~\eqref{eq:Hgeneral}] reduces to~\cite{sharma2023}:

\begin{equation}
\begin{aligned}
    E_{\rm K} &= \frac{\tilde{K}_1}{2N} \sum_{\nu \neq \nu'} \left(\tilde{\mathbf{S}}_{{\mathbf{Q}}_{\nu}} \cdot \tilde{\mathbf{S}}_{{\mathbf{Q}}_{\nu'}}\right)^2
+ \frac{\tilde{K}_2}{N} \sum_{\nu=1}^{3}\left(\tilde{\mathbf{S}}_{{\mathbf{Q}}_{\nu}} \cdot \tilde{\mathbf{S}}_{{\mathbf{Q}}_{\nu}}\right)^2
\\
&+ \frac{\tilde{K}_3}{2 N} \sum_{\nu \neq \nu'} \left(\tilde{\mathbf{S}}_{{\mathbf{Q}}_{\nu}} \cdot \tilde{\mathbf{S}}_{{\mathbf{Q}}_{\nu}}\right) \left(\tilde{\mathbf{S}}_{{\mathbf{Q}}_{\nu'}} \cdot \tilde{\mathbf{S}}_{{\mathbf{Q}}_{\nu'}}\right) 
\end{aligned}
\label{H_4spin}
\end{equation}

\begin{figure*}[ht]
  \includegraphics[width=1\linewidth]{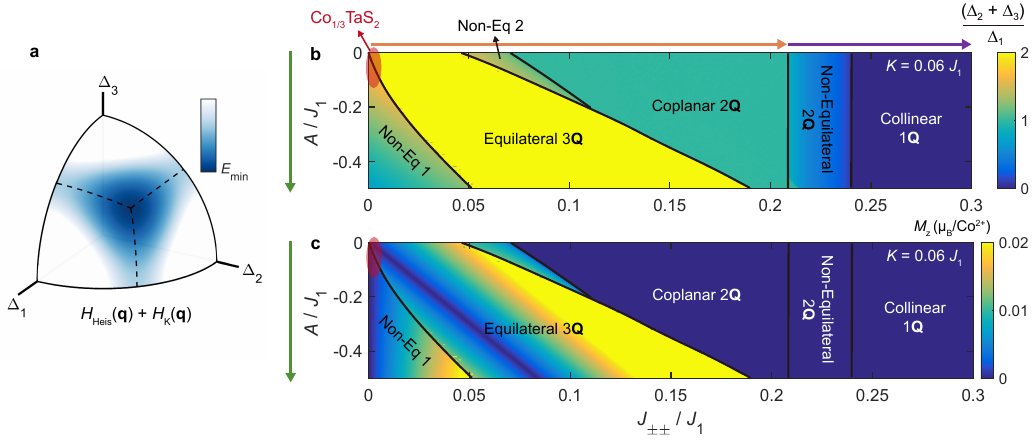}%\\
  \caption{Theoretical zero-temperature phase diagram of the isotropic and anisotropic spin Hamiltonian. \textbf{a}, Classical energy landscape of the isotropic spin Hamiltonian (Eq.~\eqref{eq:Hgeneral}), indicating an equilateral triple-$\mathbf{Q}$ ground state for $\tilde{K}>0$. Note that the Heisenberg term yields a constant energy across the entire variation space, making the landscape solely dependent on the four-spin interaction term (Eq.~\eqref{E_4spin_simplified}). \textbf{b}--\textbf{c}, Phase diagram spanned by single-ion anisotropy $A$ and bond-dependent exchange anisotropy $J_{\pm\pm}$, in the presence of $K=0.06J_{1}$. The colour scales in \textbf{b} and \textbf{c} represent $\frac{\Delta_{2} + \Delta_{3}}{\Delta_{1}}$ ($\Delta_{3} \leq  \Delta_{2} \leq \Delta_{1}$) and the out-of-plane net magnetization ($M_{z}$), respectively, as obtained from our ground state search algorithm (see Section II. E of Supplementary Note). The spin configurations of each phase are shown in Fig.~\ref{Sfig:multiQ}c and Fig.~\ref{Sfig:Fourier}.}
\label{Sfig:multiQpdg}
\end{figure*}

The first term vanishes ($E_1 =0$) because of the orthogonality condition given in Eq.~\eqref{eq:ortho}. In addition, Eq.~\eqref{eq:sumrule} implies that $E_2$ and $E_3$ are connected by a simple linear relation
\begin{eqnarray}
E_2 = -2 E_3 + {N}S^{4},
\label{E2_E3}
\end{eqnarray}
implying that  $\langle \cal H_{\rm K}\rangle$ depends on a single coefficient:
\begin{eqnarray}
E_{\rm K} =
 (\tilde{K}_{2}-\frac{\tilde{K}_{3}}{2})E_2 + \frac{\tilde{K}_{3}S^{4}}{2} \equiv
 \tilde{K}  E_2 + \mathrm{const},
\label{E_4spin_simplified}
\end{eqnarray}
with
\begin{eqnarray}
E_2  &=& \frac{1}{N} \langle \sum_{\nu=1}^{3}{\left(\tilde{\mathbf{S}}_{{\mathbf{Q}}_{\nu}} \cdot \tilde{\mathbf{S}}_{{\mathbf{Q}}_{\nu}}\right)}^2 \rangle = \frac{1}{N}( {\Delta_{1}}^4+{\Delta_{2}}^4+{\Delta_{3}}^4).
\nonumber \\
\end{eqnarray}
Consequently, the combination of $\cal H_{\rm Heis}$ and $\cal H_{\rm K}$ can produce only two possible magnetic ground states: (1) an equilateral triple-$\mathbf{Q}$ magnetic ordering (${\Delta_{1}}={\Delta_{2}}={\Delta_{3}}$) when $\tilde{K}>0$, and (2) a stripe single-$\mathbf{Q}$ ordering ($\Delta_{1}=\sqrt{N}S$ and ${\Delta_{2}}={\Delta_{3}}=0$) when $\tilde{K}<0$. The colour plot in Fig.~\ref{Sfig:multiQpdg}a shows $E_2$ in the ${\Delta_{1}}-{\Delta_{2}}-{\Delta_{3}}$ space, illustrating the energy landscape for $\tilde{K}>0$, which indeed favors the equilateral configuration where ${\Delta_{1}}={\Delta_{2}}={\Delta_{3}}$.

In \CTS{}, $\tilde{K}>0$ accounts for the triple-$\mathbf{Q}$ and single-$\mathbf{Q}$ nature of the magnetic orderings in \Tb{} (Phase I) and \Ti{} (Phase III), respectively \cite{park2023_ncomm, park2024_INS}. Although $\tilde{K}>0$ leads to a triple-$\mathbf{Q}$ ground state at zero temperature, collinear single-${\mathbf{Q}}$ ordering can still emerge as a finite-temperature ground state due to the order-by-thermal-disorder mechanism \cite{od_by_disod1, od_by_disod2}. However, the equilateral ground state predicted by this generalized four-spin model fails to account for the broken three-fold rotational symmetry ($C_{3z}$) observed in \Tb{} and $H=0$ through our MLD measurements. This discrepancy motivates the inclusion of magnetic anisotropy in our model, which is supported by several experimental observations in \CTS{} (see Section II. D).

\subsection{Equivalence to the real-space scalar biquadratic interaction model}
Before including magnetic anisotropy, we discuss the correspondence between the Fourier transformed $\hat{\cal H}_{\rm K}$, simplified into Eq.~\eqref{E_4spin_simplified}, and the simplest real-space scalar biquadratic interaction term ($\mathcal{\hat H}_{\rm bq}$) between nearest-neighbors (NNs):

\begin{eqnarray}
\mathcal{\hat H}_{\rm bq} &=& {K}\sum_{{\mathbf{r}}, {\bm{\delta}}_1} (\hat {\mathbf{S}}_{\mathbf{r}} \cdot \hat {\mathbf{S}}_{{\mathbf{r}}+{\bm{\delta}}_1})^{2},
\label{4spin_rspace}
\end{eqnarray}
where $\bm\delta_{1}$ runs over the bond vectors connecting NNs in a triangular lattice, without double-counting. As far as the bilinear Heisenberg interactions are much stronger than $K$ and produce $M$-ordering, the most general magnetic ordering is still a  four-sublattice structure, as dictated by Eq.~\eqref{eq:transform}. An explicit form of Eq.~\eqref{4spin_rspace} based on these four spin sublattices is obtained as follows:
\begin{eqnarray}
E_{\rm bq} = \frac{{K}}{N} \sum_{j \neq j'} (\mathbf{S}_j\cdot\mathbf{S}_{j'})^2
\label{Kbq_4spin}
\end{eqnarray}
with $j,j'=1,2,3,4$.

Replacing $\mathbf{S}_{i}$ in Eq.~\eqref{Kbq_4spin} with $\tilde{\mathbf{S}}_{\mathbf{Q}_{\nu}}$ (assuming $|\tilde{\mathbf{S}}_{\mathbf{0}}| \ll S$) yields:

\begin{eqnarray}
E_{\rm bq} = \frac{2{K}}{N}\Big[&6&\sum_{\nu=1}^{3} \Delta_{\nu}^4 - \sum_{\nu \neq \nu'}
\Delta_{\nu}^2 \Delta_{\nu '}^2\Big].
\end{eqnarray}
Using Eq.~\eqref{eq:sumrule}, it can be shown that the sum of the last three terms is proportional to the first term, up to an additive constant. This results in the following simplified expression:
\begin{eqnarray}
E_{\rm bq} = 14 \frac{K}{N} \Big[ \sum_{\nu=1}^{3}{{\Delta^4_{\nu}}} -\frac{N^2S^{4}}{7}\Big] 
= 14K E_{2} + \mathrm{const}.
\nonumber \\
\label{Kbq_converted}
\end{eqnarray}
The $E_{\rm bq}$ term precisely coincides with the simplified form of the Fourier transformed $E_{\rm K}$ derived in the previous section, except for a different constant and scaling factor, which do not alter the overall energy landscape. Thus, $E_{\rm bq}$ produces the same classical energy landscape as $E_{\rm K}$ shown in Fig.~\ref{Sfig:multiQpdg}a, implying that a bi-quadratic interaction is enough to model the most general classical ground state of the 4-sublattice structure and its long-wavelength (i.e., low-energy) fluctuations. 

This equivalence is particularly valuable for numerical spin simulations. In \CTS{}, incorporating thermal fluctuations is essential for realizing Phase III or Phase IV, for which classical Monte Carlo simulations have proven highly effective \cite{park2023_ncomm, park2024_INS}. However, classical Monte Carlo simulations can only be performed with a real-space spin Hamiltonian. The equivalence demonstrated in this subsection ensures that we can accurately simulate the low-energy physics of the generalized momentum-space Hamiltonian using the real-space Hamiltonian, which includes Heisenberg interaction terms and the nearest-neighbor (NN) scalar bi-quadratic term $\mathcal{\hat H}_{\rm bq}$. However, while this model can describe the ground state and low-energy modes, it should not be considered a microscopic Hamiltonian for \CTS{}. The underlying physics of long-range exchange interactions in metallic \CTS{} is more accurately captured by the Fourier-transformed form $\hat{\cal H}_{\rm K}$, which avoids introducing any artificial cut-off in the real-space exchange interactions.

The effective bilinear and biquadratic exchange constants of the real-space spin Hamiltonian were determined in previous inelastic neutron scattering (INS) studies. Notably, Ref.~\cite{park2024_INS} proposed a reliable model that accounts for both intralayer and interlayer couplings up to a bond length of 11.5\,\AA. This model successfully describes the low-energy spin dynamics of \CTS{} in both Phase I and Phase III, as well as the full spin dynamics in the paramagnetic phase (\Th{}). All classical Monte Carlo simulations and zero-temperature phase diagrams were computed using the bilinear exchange coefficients and $K$ suggested in Ref.~\cite{park2024_INS}, along with the magnetic anisotropy terms introduced in Section II. D below.

\subsection{Magnetic anisotropy}

In this section, we introduce two types of magnetic anisotropy present in \CTS{}, which are crucial for explaining our observations from magneto-optical measurements. Generally, easy-axis magnetic anisotropy favors collinear single-$\mathbf{Q}$ magnetic ordering, as shown in Figs.~\ref{fig:theory}e and ~\ref{Sfig:multiQ}c. When the anisotropy is much weaker than the four-spin interactions, the system retains the equilateral triple-$\mathbf{Q}$ ordering described in Section II. B. However, when both perturbations are comparable—our focus in this work—the competition between them can result in intermediate states that interpolate between the equilateral triple-$\mathbf{Q}$ and single-$\mathbf{Q}$ orderings. We demonstrate that this competition can lead to a non-equilateral triple-$\mathbf{Q}$ ground state, which is consistent with the symmetry and chirality features of Phase I suggested by our MLD and MCD measurements.

The primary term of interest is the single-ion easy-axis anisotropy:
\begin{eqnarray}
\mathcal{\hat H}_{\rm SI} = A \sum_{\mathbf{r}}  (\hat{S}^{z, a}_{\mathbf{r}})^2,
\label{Eq:Kea}
\end{eqnarray}
where $A <0$ and $S^{z,a}_{\mathbf{r}}$ is the out-of-plane component of the spin vector at unit cell $\mathbf{r}$ and sublattice $a$. Phase III in Fig.~2c suggests the presence of this term. Previous neutron diffraction studies have shown that this single-$\mathbf{Q}$ state comprises only out-of-plane components \cite{park2023_ncomm, takagi2023} (see Fig.~\ref{fig:theory}e), indicating that the $c$-axis is the easy-axis for the magnetic moments in \CTS{}. This anisotropy can alternatively be described by introducing XXZ exchange anisotropy in the bilinear interactions, which produces the same effect as $\mathcal{\hat H}_{\rm SI}$. 

The second source of magnetic anisotropy is $J_{\pm\pm}$, which, along with $J_{z\pm}$, represents the symmetry-allowed bond-dependent anisotropic exchange terms (see Ref. \cite{generalized_TLAF}):
\begin{eqnarray}
{\cal \hat{H}_{\mathrm{\pm\pm}}} = \sum_{{\langle i,j \rangle}_{1}} 2J_{\pm\pm} \Big[ \left( S_i^x S_j^x - S_i^y S_j^y \right) \cos \phi_\alpha \nonumber \\ 
- \left( S_i^x S_j^y + S_i^y S_j^x \right) \sin \phi_\alpha \Big]
\label{Eq:Jpmpm}
\end{eqnarray}
where ${{\langle i,j \rangle}_{1}}$ runs over the bonds between nearest neighbours, $x$ $\parallel$ $a$-axis, and $\phi_{\alpha} \in \{0, 2\pi/3, 4\pi/3\}$ is the angle between a bond vector ($i \to j$) and the $a$-axis (i.e., bond-dependent). Similar to $\mathcal{\hat H}_{\rm SI}$, the presence of a finite $J_{\pm\pm} > 0$ is supported by previous experimental observations. First, the $J_{\pm\pm} (> 0)$ term aligns all four spins along high-symmetry crystalline directions by breaking continuous spin-rotational symmetry~\cite{nickel2023}, as suggested by Rietveld refinement analysis from previous neutron diffraction studies \cite{park2023_ncomm,takagi2023}. This broken spin-rotational symmetry accounts for the small energy gap of the Goldstone magnon mode in the triple-$\mathbf{Q}$ phase of \CTS{} \cite{park2023_ncomm}. Second, it induces a slight out-of-plane spin canting in tetrahedral triple-$\mathbf{Q}$ orderings (both equilateral and non-equilateral), which results in a tiny net magnetization along the $c$-axis (see Fig. \ref{Sfig:multiQpdg}c), consistent with experimental observations (Fig. \ref{fig:intro}d).

It is worth noting that such anisotropy is common in triangular lattice magnets \cite{hayami2021}. However, a detailed investigation of the impact of these anisotropic terms on the $M$-ordering problem ($\mathbf{Q}_{\nu} = \mathbf{G}_{\nu}/2$) has yet to be conducted.

While accurately determining the magnitudes of $A$ and $J_{\pm\pm}$ is not feasible with the available data, their upper limits can be reasonably estimated based on previous INS studies \cite{park2023_ncomm,park2024_INS}. The direct consequence of finite $J_{\pm\pm}$ and $A$ is the opening of an energy gap in the Goldstone magnon mode for the triple-$\mathbf{Q}$ (\Tb{}) and single-$\mathbf{Q}$ (\Ti{}) phases, respectively. In the triple-$\mathbf{Q}$ phase, a small energy gap of approximately 0.5\,meV has been observed~\cite{park2023_ncomm}. A simple spin-wave calculation, including proper resolution convolution effects, suggests that for $J_{\pm\pm}$ values larger than 0.0024\,meV (approximately $0.002J_{1}$), a gap larger than 0.5\,meV is opened. Therefore, a reasonable range for $J_{\pm\pm}$ is from 0 to roughly $0.002J_{1}$.

Estimating $A$ is more complex. Experimentally, no energy gap has been detected in the single-$\mathbf{Q}$ spin-wave spectrum of \Ti{} (Phase III), with a precision down to 0.2\,meV \cite{park2024_INS}. However, it should be noted that this intermediate single-$\mathbf{Q}$ ordering emerges under significant thermal fluctuations via the order-by-thermal-disorder mechanism~\cite{od_by_disod1,od_by_disod2}. As a result, the observed spectrum should not exhibit an energy gap for small enough values of $|A|$. Our preliminary investigation into the relationship between $|A|$ and the energy gap under significant thermal fluctuations, using Landau-Lifshitz dynamics (LLD) simulations within the Su(n)ny software package~\cite{sunny_mainref,Sunny}, suggests that the approximate range for $A$ in \CTS{} is between 0 and $0.06J_{1}$, which is smaller than $|K|$.

\subsection{Phase diagram spanned by anisotropy}

In this section, we explain how magnetic anisotropy naturally accounts for the rich phase diagram observed in MLD and MCD measurements  (Fig.~\ref{fig:hysttemperature}c). Notably, it accurately predicts Phase I by producing a non-equilateral triple-$\mathbf{Q}$ ground state, which breaks three-fold rotational symmetry while preserving nearly the same chiral structure as the equilateral triple-$\mathbf{Q}$ ordering.

\begin{figure}[t]
  \includegraphics[width=1\columnwidth]{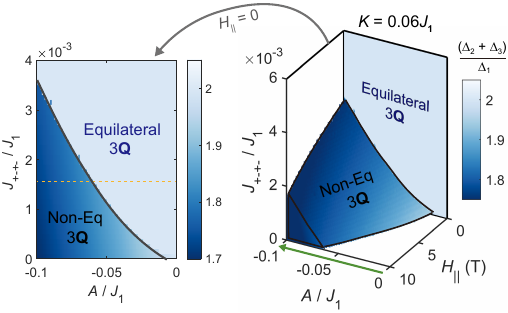}%\\
  \caption{Theoretical zero-temperature phase diagram of the anisotropic spin Hamiltonian for \CTS{} (Eqs.~\eqref{Eq:Kea} and \eqref{Eq:Jpmpm}) under an out-of-plane magnetic field ($H$). While a non-zero $A$ induces the non-equilateral (non-Eq) 3$\mathbf{Q}$ state, $H$ favors the equilateral 3$\mathbf{Q}$ state. The left panel shows the $H=0$ slice of the $A-J_{\pm\pm}-H$ phase diagram.}
\label{Sfig:3Dpdg}
\end{figure}

We calculated the $T=0$ magnetic phase diagram for the spin Hamiltonian with isotropic bilinear and biquadratic interactions, as well as $\mathcal{\hat H}_{\rm SI}$ and ${\cal \hat{H}_{\mathrm{\pm\pm}}}$ discussed above. This was achieved by minimizing the classical energy of the four spin sublattice configuration $\mathbf{S}_{j}~(j=1,2,3,4)$ using the conjugate gradient method implemented in the Su(n)ny software package~\cite{sunny_mainref,Sunny}. The four vector Fourier components in Eq.~\ref{eq:transform} were subsequently calculated from the optimal real space configuration $\mathbf{S}_{j}$. From these components, we can classify the ground state orderings as: single-$\mathbf{Q}$ (one non-zero $\Delta_{1,2,3}$), double-$\mathbf{Q}$ (two non-zero $\Delta_{1,2,3}$), non-equilateral triple-$\mathbf{Q}$ (three non-zero $\Delta_{1,2,3}$ but with different magnitudes), and equilateral triple-$\mathbf{Q}$ (three non-zero $\Delta_{1,2,3}$ with equal magnitudes). These states are effectively visualized using the ratio $(\Delta_{2}+\Delta_{3})/\Delta_{1}$ where $\Delta_{1}>\Delta_{2}>\Delta_{3}$. 

\begin{figure}[h!]
  \includegraphics[width=0.88\columnwidth]{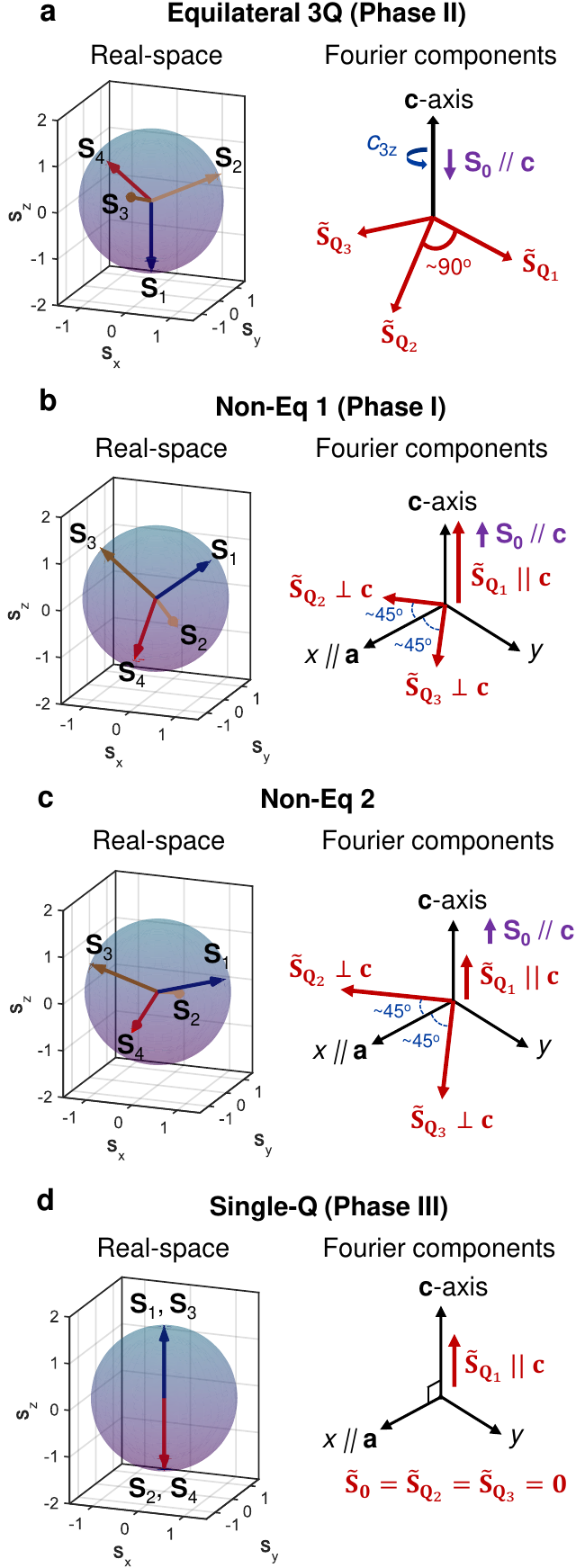}%\\
  \caption{Real-space configurations of the four spin sublattices for several multi-$\mathbf{Q}$ magnetic ground states predicted by our anisotropic spin Hamiltonian; see Fig.~\ref{Sfig:multiQpdg}b. The right side of each panel illustrates the corresponding vector Fourier components $\tilde{\mathbf{S}}_{\mathbf{Q}_{\nu}}$ ($\nu$=1,2,3).}
\label{Sfig:Fourier}
\end{figure}

Fig.~\ref{Sfig:multiQpdg}b presents the resultant zero-temperature magnetic phase diagram over a wide range of $A$ and $J_{\pm\pm}$ values, in units of $J_{1}=1.212\,$meV \cite{park2024_INS}. The isotropic interaction coefficients were adopted from Ref.~\cite{park2024_INS}: $J_{1}=1.212\,$meV, $J_{2}=0.264J_{1}$, $J_{3}=0.018J_{1}$, $J_{c1}=1.160J_{1}$, $J_{c2}=-0.215J_{1}$, and $K=0.06J_{1}$, where $J_{n}$ ($J_{cm}$) denotes the interaction between $n^{\mathrm{th}}$ intralayer ($m^{\mathrm{th}}$ interlayer) NN spins. As described in the previous section, increasing the strength of each anisotropy gradually transforms the equilateral triple-$\mathbf{Q}$ ground state into a single-$\mathbf{Q}$ state, with a non-equilateral triple-$\mathbf{Q}$ ground state emerging in the intermediate regime. Notably, increasing $|A|$ and $J_{\pm\pm}$ follows two distinct high-symmetry routes on the continuous multi-$\mathbf{Q}$ manifold. The path driven by $A$ (shown by green arrows) and the path driven by $J_{\pm\pm}$ (shown by orange and purple arrows) are illustrated throughout Fig.~\ref{Sfig:multiQ}b--c and ~\ref{Sfig:multiQpdg}b.

Based on the feasible magnitudes of $|A|$ and $J_{\pm\pm}$ discussed in the previous section, we have indicated the location of \CTS{} on the phase diagram (shown as a red ellipsoidal region in Fig.~\ref{Sfig:multiQpdg}b). An enlarged view of this region is displayed in the left panel of Fig.~\ref{Sfig:3Dpdg}. Notably, once $A$ becomes non-zero, the ground state immediately transitions from the equilateral triple-$\mathbf{Q}$ state to the non-equilateral triple-$\mathbf{Q}$ state (Non-Eq1 in Fig.~\ref{Sfig:multiQ}c). Thus, it is natural for a system with slight easy-axis anisotropy to exhibit non-equilateral triple-$\mathbf{Q}$ ordering with broken three-fold rotational symmetry, which is consistent with our observation of Phase I.

Fig.~\ref{Sfig:Fourier}b shows the real-space spin configuration and the corresponding Fourier components of the Non-Eq1 phase, which we identify as the most likely magnetic structure for Phase I in Fig.~2c. First, the orientations of the three Fourier components differ from those of the three-fold symmetric equilateral ordering (Fig.~\ref{Sfig:Fourier}a): while $\tilde{\mathbf{S}}_{\mathbf{Q}_{1}}$ is parallel to the $c$-axis, $\tilde{\mathbf{S}}_{\mathbf{Q}_{2}}$ and $\tilde{\mathbf{S}}_{\mathbf{Q}_{3}}$ lie within the $a-b$ plane and are separated by $\pm(45^{\circ} + \delta)$ from the $a$-axis, where $\delta \ll 1$ depends on the magnitude of the weak ferromagnetic moment along the $c$-axis (see Fig.~\ref{Sfig:multiQpdg}c). Second, while the two in-plane Fourier components have equal magnitudes, the out-of-plane Fourier component is larger ($\Delta_{1} > \Delta_{2} = \Delta_{3}$), placing this phase as an intermediate state between the equilateral triple-$\mathbf{Q}$ and single-$\mathbf{Q}$ orderings. Notably, both of these features break the three-fold rotational symmetry in this phase (i.e., the symmetry is reduced to $C_{2z}$).

For comparison, we also present the Non-Eq2 phase in Fig.~\ref{Sfig:Fourier}c. Although this phase shares the same vector Fourier component configuration as the Non-Eq1 phase (Fig.~\ref{Sfig:Fourier}b), the two in-plane Fourier components have a greater magnitude than the out-of-plane component ($\Delta_{1} < \Delta_{2} = \Delta_{3}$), indicating its nature as an intermediate phase between the equilateral triple-$\mathbf{Q}$ and double-$\mathbf{Q}$ states. However, realizing this phase would require an order-of-magnitude larger $J_{\pm\pm}$ than what was estimated based on the spin gap. Therefore, it is unlikely that this phase corresponds to Phase I in \CTS{}.

\subsection{External magnetic field}
To study the transition between Phase I and Phase II in Fig.~\ref{fig:hysttemperature}c and ~\ref{fig:theory}, we additionally incorporated the out-of-plane magnetic field into the spin model via the Zeeman term:
\begin{eqnarray}
{\cal \hat{H}}_{\mathrm{z}} =  g \mu_{\mathrm{B}} H_{\parallel} \sum_{\mathbf{r}, a} \hat{S}^{z,a}_{\mathbf{r}},
\label{Eq:Zeeman}
\end{eqnarray}
where $g=2$ (the orbital magnetic moments are believed to be quenched in \CTS{}) and $H_{\mathrm{||}}$ denotes the strength of the out-of-plane field. The full three-dimensional phase diagram, spanned by $A$, $J_{\pm\pm}$, and $H_{\mathrm{||}}$, is shown in the right panel of Fig.~\ref{Sfig:3Dpdg}. Indeed, Eq.~\eqref{Eq:Zeeman} favors the equilateral triple-$\mathbf{Q}$ magnetic ordering (Fig.~\ref{Sfig:Fourier}a) over the non-equilateral triple-$\mathbf{Q}$ state (Fig.~\ref{Sfig:Fourier}b), thereby restoring the three-fold rotation symmetry under a strong out-of-plane magnetic field. This is consistent with our observation that the MLD signal disappears in Phase II when $H>H_\textrm{m}$; see Fig.~\ref{fig:hysttemperature}c. The temperature-field phase diagram, obtained by classical Monte Carlo simulations using reasonable values for $A$ and $J_{\pm\pm}$ for \CTS{} (namely, $A=0.06J_{1}$ and $J_{\pm\pm}=0.0016J_{1}$) is shown in Fig.~\ref{fig:theory}f--g. As demonstrated, our spin model successfully captures the experimental observations related to Phase I, II, III.

\end{document}